\documentclass[a4paper,11pt]{article}
\usepackage[utf8]{inputenc}
\usepackage[english]{babel}
\usepackage{amsmath}
\usepackage{amssymb}
\usepackage{dsfont}
\usepackage{jheppub}
\usepackage{physics}
\usepackage{color}
\usepackage{etaremune}
\usepackage{graphicx}
\usepackage{comment}
\usepackage[multiple]{footmisc}
\usepackage{mathtools}
\usepackage{bm}
\usepackage{amsmath,amssymb,amsfonts,mathrsfs}

\usepackage{hyperref}
\hypersetup{colorlinks=true,linkcolor=blue,citecolor=red,urlcolor=black,pdfauthor={Giulio Neri}}

\newcommand{\SISSA}{\affiliation{SISSA, International School for Advanced Studies, \\
via Bonomea 265, 34136 Trieste, Italy}}
\newcommand{\InfnTS}{\affiliation{INFN, Sezione di Trieste,\\
via Valerio 2, 34127 Trieste, Italy}}
\newcommand{\IFPU}{\affiliation{IFPU, Institute for Fundamental Physics of the Universe,\\
via Beirut 2, 34014 Trieste, Italy}}

\DeclarePairedDelimiter\floor{\lfloor}{\rfloor}
\def\dbar{{\mkern3mu\mathchar'26\mkern-12mu \delta}}

\title{Covariant phase space analysis \\ of Lanczos-Lovelock gravity with boundaries}
\author{Giulio Neri} \author{and Stefano Liberati}

\SISSA\IFPU\InfnTS

\emailAdd{gneri@sissa.it}
\emailAdd{liberati@sissa.it}
\date{\today}

\abstract{This work introduces a novel prescription for the expression of the thermodynamic potentials associated with the couplings of a Lanczos-Lovelock theory. These potentials emerge in theories with multiple couplings, where the ratio between them provide intrinsic length scales that break scale invariance. Our prescription, derived from the covariant phase space formalism, differs from previous approaches by enabling the construction of finite potentials without reference to any background. To do so, we consistently work with finite-size systems with Dirichlet boundary conditions and rigorously take into account boundary and corner terms: including these terms is found to be crucial for relaxing the integrability conditions for phase space quantities that were required in previous works. We apply this prescription to the first law of (extended) thermodynamics for stationary black holes, and derive a version of the Smarr formula that holds for static black holes with arbitrary asymptotic behaviour.}
\arxivnumber{}

\begin{document}

\maketitle

\section{Introduction}
\label{ss.Introduction}

Although the profound connection between gravity and thermodynamics was first discovered in the early 70s~\cite{Hawking:1975vcx, Bekenstein:1973ur, Bardeen:1973gs}, its ongoing influence on contemporary theoretical physics research cannot be overstated.
The reason for such an interest is that, for classical physics, the link between these two aspects appears paradoxical, while it starts to make sense once the quantum nature of fields is taken into account. Therefore, progress in understanding this connection has been considered relevant towards our our understanding of how gravity and quantum physics fit together.

However, the fundamental clash between the formalism of General Relativity (GR) and thermodynamics hindered early attempts. In fact, the latter is naturally built upon the Hamiltonian formulation of mechanics, which involves the identification of a preferred time direction, whereas the former is based on a very powerful symmetry principle, the one of general covariance, which one would like to keep manifest as much as possible throughout the calculations.
It is not surprising then that a pivotal step in this field of research was the introduction of the \textit{covariant phase space formalism}, which allowed to formulate Hamiltonian mechanics while preserving explicit covariance~\cite{Iyer:1994ys, Iyer:1995kg, Lee:1990nz, Wald:1993nt, Wald:1984rg, Wald:1999wa}.
After that, the formalism has found widespread application in many fields, ranging from AdS/CFT~\cite{Hollands:2005wt, Compere:2008us, Andrade:2015fna, Belin:2018bpg, Belin:2018fxe} to near-horizon symmetries~\cite{Carlip:1999cy, Haco:2018ske, Haco:2019ggi}, from asymptotic symmetries~\cite{Barnich:2001jy, Barnich:2011mi, Strominger:2017zoo} to non-stationary black holes~\cite{Wall:2015raa, Wall:2024lbd, Visser:2024pwz, Hollands:2024vbe}.

The covariant phase space formalism {is especially valuable for theories that are generally covariant, such as those describing gravity, as most famously exemplified by GR} (see e.g.~\cite{Faraoni:2010,Hajian:2020dcq,Heisenberg:2022nvs,Minamitsuji:2023nvh}). Over the last century, {many theories of gravity have been developed to extend GR}, and most of them\,---\,if not all\,---\,contain black holes. These objects, despite their phenomenological differences, appear to obey the same laws of thermodynamics 
\cite{Pacilio:2017swi,Cai:2003kt,Kubiznak:2016qmn,Minamitsuji:2023nvh}. {In this work, we will focus our attention only on the first law.}

A powerful formula that one can deduce from the first law of black hole thermodynamics is the Smarr formula~\cite{Smarr:1972kt}, which relates the black hole mass to its other macroscopic properties, like its horizon area, surface gravity and angular momentum. The relevance of this formula is that it provides a relation that can be used also when the analytic form of the solution is unknown. In GR, the Smarr formula looks very simple
\begin{equation}
    M=2\,\Omega_H J+\,\frac{\kappa A_H}{4\pi G_N}.
\end{equation}
However, this formula does not hold as soon as spacetime has a non-zero cosmological constant. Even worse, it appears that no particularly simple relation can be derived from the first law. This problem has been noticed long ago~\cite{Magnon:1985sc} and it is related to the breaking of scale invariance by the cosmological constant~\cite{Caldarelli:1999xj} (a similar problem appeared in~\cite{Rasheed:1997ns}). 

The same issue appears more generally in the context of Lanczos-Lovelock (LL) theories in more than four dimensions. These theories are characterized by the presence of higher curvature Lagrangians which, despite the appearance, are devised in such a way to lead to second order field equations (as we shall see in detail in Sec.~\ref{s.llt}). A generic LL theory is given by a sum of such Lagrangian terms with arbitrary couplings. However, whenever two or more of such couplings are non-zero, the recovery of the Smarr formula appears to be problematic as in the case of GR with a cosmological constant.

In all the above mentioned cases, what one finds is that the mass depends on the (ratio between) couplings trough some thermodynamic potentials~\cite{Kastor:2008xb, Kastor:2009wy, Kastor:2010gq,Liberati:2015xcp}.
Such potentials quantify the variation of the total Noether charge of the solution with respect to a change in the (relative) value of the couplings of the theory, as first suggested in~\cite{Gibbons:2004ai} and confirmed in~\cite{Wang:2006bn}.
In this work, we provide a simple derivation of these potentials and, consequently, of the Smarr formula, specializing our results to LL theories. We hope that streamlining the construction of the potentials makes it easier to extract their physical interpretation.

Let us finally add that another problem of the first investigations in covariant phase space was the lack of rigor in the treatment of boundary effects. For example, in order to define conserved charges from their action on phase space, most authors needed to assume the existence of a certain boundary quantity to integrate the variation of the symplectic form. However, the role of boundaries in the covariant phase space formalism has been shown to be crucial to provide clear prescription for the charges.
Recently, the authors of Ref.~\cite{Harlow:2019yfa} carried out a careful derivation taking into account boundary effects for many theories, including GR. Our work aims to extend such a rigorous treatment to LL theories, where the aforementioned presence of extra couplings have so far hampered a complete understanding of black hole thermodynamics.

This paper is organized as follows. We devote Section \ref{s.cpsf} to the presentation of the covariant phase space formalism. Here, we establish the notation and language that will be used throughout the paper. In Section \ref{s.llt}, we review Lanczos-Lovelock theories in a form suitable for the covariant phase space treatment. In the last part of the section, we introduce the first law of black hole thermodynamics in its differential form, derived from the covariant phase space algorithm. We also generalize it to variations that violate boundary conditions. The central and most important part of the paper is Section \ref{s.bhs}, since we apply here the covariant phase space machinery to the LL theories in order to derive (in a straightforward way) the Smarr formula. We begin by considering Schwarzschild-like solutions of LL theories, namely, static and spherically symmetric spacetimes containing a black hole. Next, we review and analyze the issues arising in constructing the Smarr formula in this class of theories, emphasizing the significance of on-shell scale invariance breaking. We conclude the section by presenting two derivations of the Smarr formula, which involve a novel method for computing the thermodynamic potentials conjugated to the LL couplings.
In the final section (\ref{s.ced}), we provide a thermodynamic description of these solutions within the framework of Euclidean (quantum) gravity to justify and test our theoretical construction.

Our convention is to work with the ``mostly plus'' signature $(-,+,+,\dots)$, so that the norm of timelike vectors is defined as $\norm{v}\equiv\sqrt{-v^2}$. We use greek letters to write indices of spacetime tensors, and latin letters to write indices of boundary tensors (with the only exception of $i$ and $k$ that we use also use as indices of sums).

\section{Covariant phase space formalism}
\label{s.cpsf}

We start by briefly reviewing the salient aspects of the covariant phase formalism, adhering to its modern formulation based on the (relative) variational bicomplex~\cite{Ian-M-Anderson:1989, Margalef-Bentabol:2020teu} which has been extremely valuable in our understanding of gravity~\cite{Khavkine:2014kya, Harlow:2019yfa}

\begin{figure}
    \centering
    \includegraphics[width=0.5\textwidth]{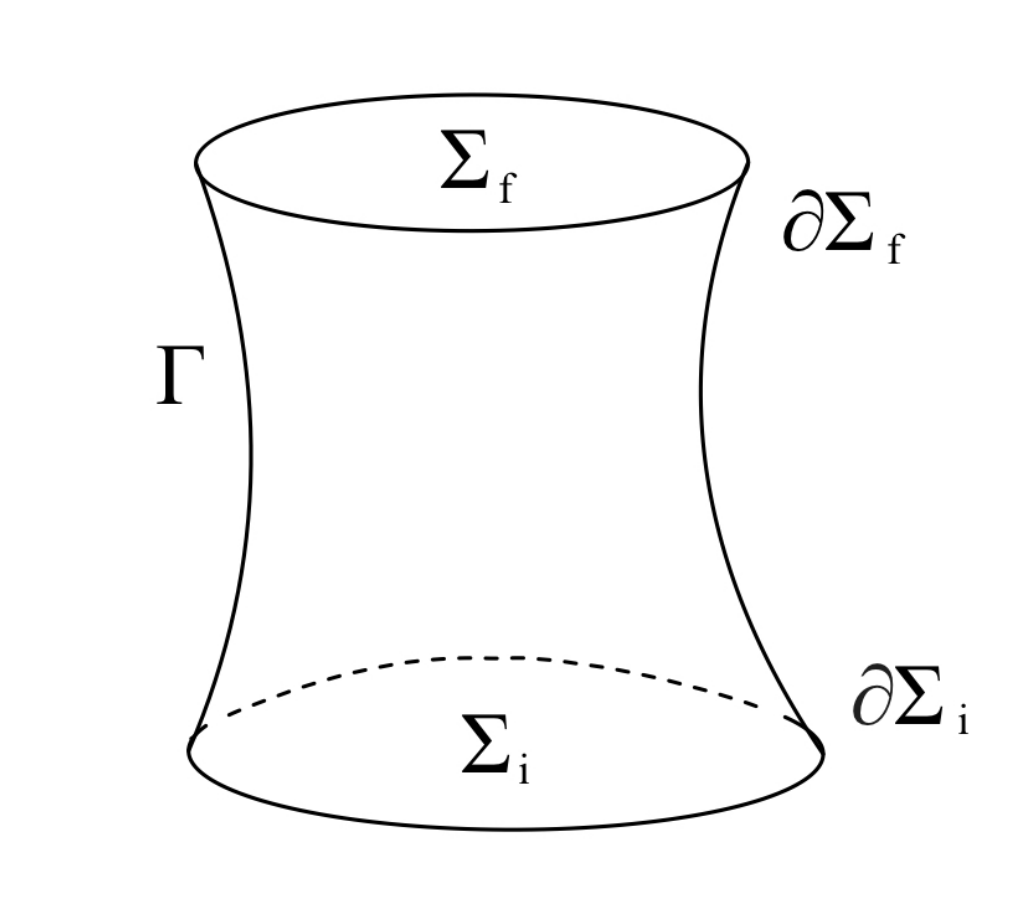}
    \caption{Representation of the cylindrical spacetime under consideration with its different parts: the lateral boundary $\Gamma$, the two ``lids'' $\Sigma_i$ and $\Sigma_f$, and the corners $\partial\Sigma_i \cup \partial\Sigma_f$}
    \label{fig:spacetime}
\end{figure}

In this work, we consider spacetime to be a $d$-dimensional, connected and oriented manifold $\mathcal M$ with a simple topology. We assume $\mathcal M$ to admit a foliation by Cauchy hypersurfaces, so that $\mathcal M=\Sigma\times[t_i,t_f]$. The boundary of $\mathcal M$ thus comprises three parts: a lateral boundary $\partial\Sigma\times[t_i,t_f]$, that we call $\Gamma$, and two ``lids'' $\Sigma_i$ and $\Sigma_f$. We emphasize that this manifold has corners $\partial\Sigma_i\cup\partial\Sigma_f$, whose presence will play a role in the proper treatment of boundary terms. This setup is depicted in Figure \ref{fig:spacetime}.

Even if we will only deal with the metric tensor field $g$, we prefer to present the formalism in a very general form here, considering a collection of fields that we denote as $\phi$. The space of all field configurations, with possible constraints (boundary conditions) at the lateral boundary $\Gamma$ is called \textit{configuration space} $\mathcal C$. The space of on-shell configurations is obtained as the intersection of $\mathcal C$ with the general set of solutions to the field equations $\Pi$ and we denote it as $\mathcal P$. The main insight of the formalism is to identify this space with the canonical phase space of classical mechanics~\cite{Segal:1960, Segal:1967nxa} (the idea has been explicitly put forward in~\cite{Witten:1986qs, Crnkovic:1987tz, Crnkovic&Witten87}): in a nutshell, every initial condition, which must specified on a certain Cauchy surface\,---\,whose choice breaks covariance\,---\,, corresponds to a unique (with many caveats) solution, which is instead defined in a covariant fashion by the extremization of a (local) action. The space $\mathcal P$ is then equipped with a natural \textit{symplectic structure} that can be algorithmically constructed\footnote{Barring simple cases, the structure which follows from the covariant phase space algorithm is not strictly symplectic because there might be vector fields that have vanishing product with any other. This situation is ever present in gauge theories, where gauge transformations induce degenerate vector fields. In those cases, the solution is to consider the quotient of $\mathcal P$ by the action of the gauge group.} and that allows us to define Poisson brackets (better known as Peierls bracket) without the need to introduce a canonical basis of configuration space variables and momenta~\cite{Peierls:1952, Peierls:1952cb}. Let us now lay out the main steps of the covariant phase space algorithm.

We start from a local field theory described by the action\footnote{We use subscripts to indicate the natural co-degree of spacetime forms (even if we take some license on this), which coincide with the codimension of the (sub-)manifold on which they are naturally integrated. For example, the Lagrangian density $\mathbf L_0$ is naturally integrated over the entire spacetime.}
\begin{equation}
\label{eq.action}
    I=\int_{\mathcal M} \mathbf{L}_0+\int_{\Gamma} \mathbf{L}_1,
\end{equation}
where the Lagrangians are local functions of $\phi$ and an arbitrary (but finite) number of its derivative.
The action is obviously invariant under the change $(\mathbf{L}_0,\mathbf{L}_1)\to (\mathbf{L}_0+\dd\bm\ell_1,\mathbf{L}_1-\bm\ell_1)$ with any $\bm\ell_1$ that vanishes on $\Sigma_{i/f}$. Actually, one might think that there is no need to introduce a boundary Lagrangian at all, as this could be eliminated by simply choosing $\bm\ell_1=\mathbf{L}_1$. However, it is generally impossible to extend the boundary Lagrangian in the bulk in a covariant fashion, so we do not allow these redefinitions.

Varying the bulk Lagrangian with respect to the fields and performing enough integration by parts, one gets
\begin{equation}
    \delta \mathbf{L}_0=\mathbf{E}_{0}(\phi,\delta\phi)+\dd\bm\Theta_1(\phi,\delta\phi) \,. 
\end{equation}
In the expression above, $\mathbf{E}_0$ does not contain any derivative of $\delta\phi$. 
The condition $\mathbf{E}_0= 0$ defines the subset $\Pi$ of the configuration space $\cal C$ corresponding to the solutions of the field equations.
The term $\bm\Theta_1$ is called \textit{boundary symplectic potential} and, for Lagrangians that contain higher than first derivatives of the field, it depends (linearly) on both $\delta\phi$ and its derivatives.
In the spirit of the variational bicomplex framework, we think of the variation $\delta$ as an external derivative, hence inducing an exterior algebra in configuration space. One can generically have $(p,q)$-forms, where $p$ and $q$ are the grading with respect to $d$ and $\delta$ respectively, but we will not use this notation and spell out the grading in the few cases we need to.

\begin{figure}
    \centering
    \includegraphics[width=0.35\linewidth]{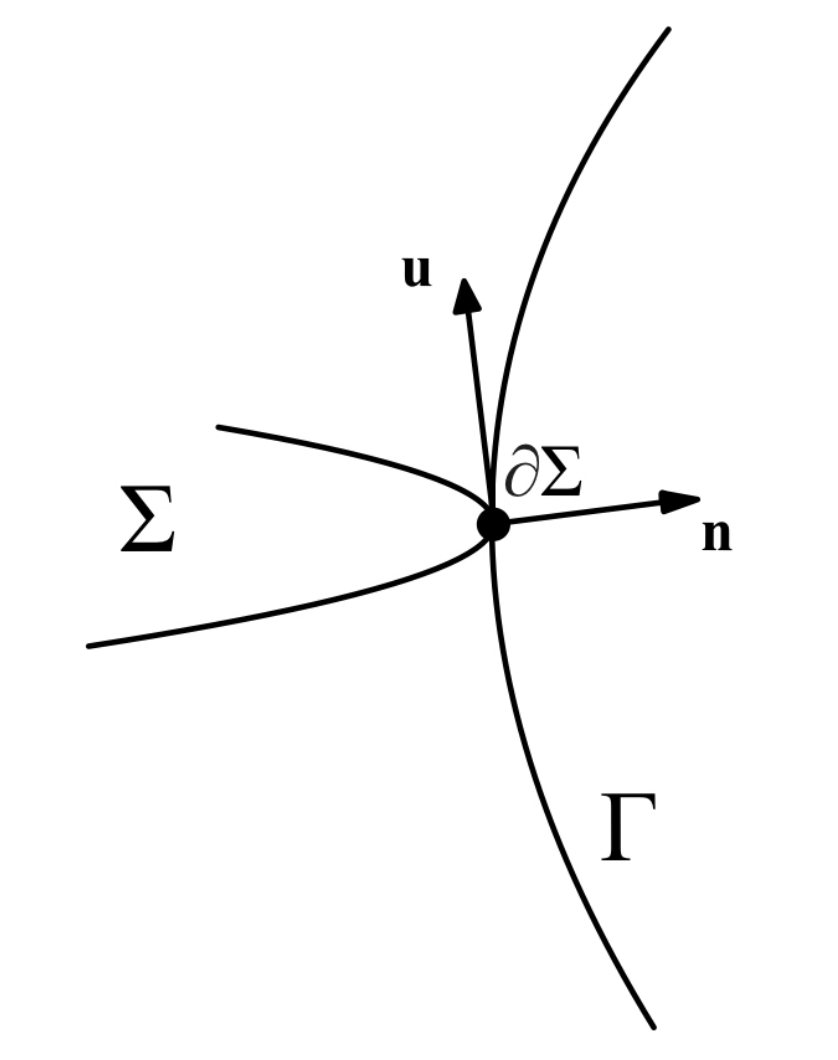}
    \caption{A representation of the two normals to the corner $\partial\Sigma$ (also referred to as $\mathcal S$ when $\Sigma$ has only one border). As we see, $\partial\Sigma$ can be seen both as the border of $\Sigma$ with normal $\mathbf{n}$ and as the border of (a subset of) $\Gamma$ with normal $\mathbf{u}$. The corresponding two induced orientations of $\partial\Sigma$ as a codimension $2$ surface are opposite.}
    \label{fig:normals}
\end{figure}

The principle of least action requires the variation of the action to vanish when evaluated for solutions of the field equations up to terms that are localized on the lids $\Sigma_i\cup\Sigma_f$. Therefore, the form of the boundary Lagrangian is not arbitrary, but it must be such that its variation cancels the terms on $\Gamma$ that one gets applying Stokes' theorem to $\dd\bm\Theta_1$. However, the cancellation need not be complete, as total derivatives will contribute only at the corner $\partial\Gamma$, which is the same set as $\partial\Sigma_i\cup\partial\Sigma_f$ and thus part of the lids~\cite{Compere:2008us, Andrade:2015gja, Compere:2011ve}. See Figure \ref{fig:normals} for reference.
We thus require the boundary Lagrangian to be such that
\begin{equation}
\label{eq.boundlagvar}
    \delta \mathbf{L}_1+j^*_\Gamma \bm\Theta_1(\phi,\delta\phi)=\mathbf{B}_{1}(\phi,\delta\phi)+\dd\bm\Theta_2(\phi,\delta\phi),
\end{equation}
where $j^*_\Gamma$ is the pull-back induced by the inclusion $j_\Gamma:\Gamma\to\mathcal M$ and $\mathbf{B}_{1}$ is required to vanish by virtue of the boundary conditions. This means that we can identify $\mathbf{B}_1=0$ as the defining condition of the configuration space $\mathcal C$ inside the space of all field configurations. Despite vanishing, we will see that the structure of $\mathbf{B}_1$ contains a lot of information and plays a relevant role when boundary conditions are allowed to vary. Finally, the term in the total derivative on the r.h.s~ of \eqref{eq.boundlagvar}, $\bm\Theta_2$, is called \textit{corner symplectic potential}.

With these conditions, the full variation of the action is\footnote{Notice that we changed the sign of $\dd\bm\Theta_2$ to account for the opposite orientation when we see the corner as $\partial\Gamma\subset\Gamma$ or as $\partial\Sigma_{i/f}\subset\Sigma_{i/f}$, as visible from Figure \ref{fig:normals}.}
\begin{equation}
    \delta I=\int_{\mathcal M}\mathbf{E}_{0}+\int_\Gamma \mathbf{B}_{1}+\int_{\Sigma_f} \pqty{\bm\Theta_1-\dd\bm\Theta_2}-\int_{\Sigma_i} \pqty{\bm\Theta_1-\dd\bm\Theta_2},
\end{equation}
so that, on the phase space $\mathcal P=\Pi\cap\mathcal C$, where both $\mathbf{E}_0=\mathbf{B}_1=0$, it has the required form
\begin{equation}
      \delta I=\int_{\Sigma_f} \pqty{\bm\Theta_1-\dd\bm\Theta_2}-\int_{\Sigma_i} \pqty{\bm\Theta_1-\dd\bm\Theta_2}.
\end{equation}
This construction singles out a particular $1$-form in configuration space $\bm\Theta_1-\dd\bm\Theta_2$. Let us notice that despite both $\bm\Theta_1$ and $\bm\Theta_2$ suffer from ambiguities, {\sl these drop out from this particular combination} \cite{Harlow:2019yfa}.
We then proceed to define the \textit{symplectic current} as the variation of the pull-back of this form on phase space
\begin{equation}
\label{eq.symcurrent}
    \bm\omega_1=\delta(\bm\Theta_1-\dd\bm\Theta_2)|_\mathcal P.
\end{equation}
This current is closed both as a $2$-form in $\mathcal P$ and as a spacetime $(d-1)$-form, and it vanishes on $\Gamma$ by the condition \eqref{eq.boundlagvar}. Finally, we construct the \textit{symplectic form} by integrating the current on a Cauchy slice $\Sigma$
\begin{equation}
    \Omega_\Sigma(\delta\phi,\delta\phi)=\int_\Sigma \bm\omega_1(\delta\phi,\delta\phi).
    \label{eq.symplF}
\end{equation}
The properties of $\bm\omega_1$ ensure that $\Omega_\Sigma$ is independent from the choice of $\Sigma$. Therefore we will drop the subscript $\Sigma$ henceforth.
The existence of a symplectic form, barring the issue of possible degeneracy, allows us to define Poisson brackets and recover the Hamiltonian formulation in the usual way.

\subsection{Diffeomorphism invariance}
\label{ss.diffinv}

{The formalism we just laid out is most useful when the theory is invariant under general coordinate transformations. We thus assume that the Lagrangian forms we work with are covariant under a certain subgroup of diffeomorphisms. Using the variational bicomplex formalism, we thus ask that, for any $\zeta$ generating a diffeomorphism in that subgroup
\begin{equation}
\label{eq.lagrangiancovariance}
    \delta_\zeta\mathbf{L}_0=\mathcal L_\zeta \mathbf{L}_0,\qquad \delta_\zeta\mathbf{L}_1=\mathcal L_\zeta \mathbf{L}_1\,,
\end{equation}
where $\delta_\zeta$ is the configuration space Lie derivative and $\mathcal L_\zeta$ the usual spacetime Lie derivative. The above expression means that the Lagrangians must transform in the same way under a spacetime diffeomorphism (they transform as forms) and under the action that the same diffeomorphism induces in configuration space. The distinction is important because symmetry transformations are allowed to act only on dynamical fields, while a diffeomorphism acts on all spacetime objects, whether they are dynamical or not.}

However, covariance of the Lagrangians is not enough. In order for a diffeomorphism to be a symmetry transformation, it must preserve both the field equations and the boundary conditions. The first requirement is realized if the total action \eqref{eq.action} is invariant. However, the presence of a boundary breaks by itself full diffeomorphism invariance, because some diffeomorphisms would ``move" the boundary. The second requirement might impose further restrictions on the allowed subgroup of transformations.

For concreteness, if we assume Dirichlet boundary conditions (this means that $\mathbf{B}_1$ vanishes when $\delta\phi=0$), then the additional requirements are
\begin{equation}
\label{eq.conditions}
    n\cdot \zeta=0,\qquad \delta_{\bar\zeta}\phi|_\Gamma=0,
\end{equation}
where $n$ is the normal to the boundary $\Gamma$ and $\bar\zeta\equiv j^*_\Gamma\zeta$, with $j^*_\Gamma$ being the pull-back to $\Gamma$.

According to Noether's theorem, any continuous symmetry of the action is associated with the existence of a current that is conserved on-shell. Let $\zeta$ be the generator of a diffeomorphism for which \eqref{eq.lagrangiancovariance} and \eqref{eq.conditions} hold. We can write the associated Noether current as
\begin{equation}
\label{eq.bulkcurrent}
\mathbf{J}_1[\zeta]=\bm{\Theta}_1(\phi,\delta_\zeta \phi)-\zeta\cdot \mathbf{L}_0,
\end{equation}
where we point out that we are evaluating $\bm\Theta_1$ along the field variation induced by the diffeomorphism and we used the dot $\cdot$ to indicate the insertion of a vector field in the first argument of the form.
Indeed, it is easy to see~\cite{Lee:1990nz} that this current is closed on-shell using Cartan's homotopy formulas and the covariance of $\mathbf{L}_0$ \eqref{eq.lagrangiancovariance}
\begin{equation}
\label{eq.bulkcurrentconservation}
    \dd \mathbf{J}_1[\zeta]=-\mathbf{E}_0(\phi,\delta_\zeta\phi)= 0.
\end{equation}

In the standard procedure, one uses this current to construct the Noether charge by integration over a spatial slice $\Sigma$. Since $\bf J_1$ is closed, the resulting charge does not vary under arbitrary bulk variations of the slice.
However, one might also want to deform the boundary $\partial\Sigma$ of the slice (still keeping it on $\Gamma$) without changing the charge. The easiest way to achieve this is to add a compensating term involving the boundary Noether current
\begin{equation}
\mathbf{J}_2[\zeta]=\bm\Theta_2(\phi,\delta_\zeta \phi)-\bar\zeta \cdot \mathbf{L}_1,
\end{equation}
whose external derivative is
\begin{equation}
    \dd\mathbf{J}_2[\zeta]=-\mathbf{B}_{1}(\phi,\delta_{\bar\zeta}\phi)+j^*_\Gamma \bm\Theta_1(\phi,\delta_{\bar\zeta}\phi).
\end{equation}
The total Noether charge can then be constructed as
\begin{equation}
\label{eq:NoetherCh}
    Q_\Sigma[\zeta]\equiv\int_\Sigma \mathbf{J}_1[\zeta]-\int_{\partial\Sigma} \mathbf{J}_2[\zeta].
\end{equation}
Let us show explicitly that evaluating the charge on different slices $\Sigma_i$ and $\Sigma_f$ makes no difference. We call $V$ the volume of spacetime bounded by $\Sigma_i \cup \Sigma_f$ and the  timelike boundary $\Gamma$. Using the inverse Stokes' theorem twice, we find
\begin{equation}
\begin{split}
\label{eq.chargeconservation}
    \Delta Q_{f/i}[\zeta]&=\int_{\Sigma_f} \mathbf{J}_1[\zeta]-\int_{\partial\Sigma_f} \mathbf{J}_2[\zeta]-\int_{\Sigma_i} \mathbf{J}_1[\zeta]+\int_{\partial\Sigma_i} \mathbf{J}_2[\zeta]=\\
    &=\int_V \dd \mathbf{J}_1[\zeta]-\int_{\Gamma} \pqty{j^*_\Gamma \mathbf{J}_1[\zeta]-\dd\mathbf{J}_2[\zeta]}=\\
    &=-\int_V \mathbf{E}_0(\phi,\delta_\zeta\phi)+\int_{\Gamma} \pqty{j^*_\Gamma (\zeta\cdot\mathbf{L}_0)-\mathbf{B}_{1}(\phi,\delta_{\bar\zeta}\phi)}.
\end{split}
\end{equation}
We then notice that $j^*_\Gamma (\zeta\cdot\mathbf{L}_0)$ is proportional to $n\cdot \zeta$, so the integral vanishes when \eqref{eq.conditions} holds and we restrict to $\mathcal P$. We thus see that charge conservation is deeply tied to the prescription of boundary conditions. Since $Q_\Sigma[\zeta]$ does not depend on $\Sigma$, we will drop the subscript henceforth.

One of the main results of the covariant phase space formalism is that, for covariant theories (with the suitable restriction on the group of diffeomorphisms that we discussed), the Noether charge $Q[\zeta]$ is an Hamiltonian function for the action that diffeomorphisms induce in the phase space $\mathcal P$ \cite{Harlow:2019yfa}. In fact, if we consider the following expressions for the bulk and boundary Noether current variations
\begin{gather}
\label{eq.bulkcurrentvariation}
    \delta \mathbf{J}_1[\zeta]=-\delta_\zeta\bm\Theta_1(\phi,\delta\phi)+\delta\bm\Theta_1(\phi,\delta_\zeta\phi)-\zeta\cdot \mathbf{E}_0+\dd(\zeta\cdot\bm\Theta_1(\phi,\delta\phi)),\\
\label{eq.boundarycurrentvariation}
    \delta \mathbf{J}_2[\zeta]=-\delta_{\bar\zeta}\bm\Theta_2(\phi,\delta\phi)+\delta\bm\Theta_2(\phi,\delta_{\bar\zeta}\phi)-\bar\zeta\cdot \mathbf{B}_1+\dd(\bar\zeta\cdot\bm\Theta_2(\phi,\delta\phi))+\bar\zeta\cdot\bm\Theta_1(\phi,\delta\phi),
\end{gather}
we readily obtain that, in phase space~\footnote{In writing $\dd\mathbf J_2[\zeta]$, we assumed that we have extended the boundary current in the bulk. The extension is however arbitrary and irrelevant since only the values of $\mathbf{J}_2[\zeta]$ on $\partial\Sigma\subset\Gamma$ actually contributes.}
\begin{equation}
\label{eq.localhamiltoneq}
    \delta(\mathbf{J}_1[\zeta]-\dd\mathbf{J}_2[\zeta])|_{\mathcal P}=-\bm{\omega}_1(\delta_\zeta \phi,\delta g)+\bm{\omega}_1(\delta \phi,\delta_\zeta \phi),
\end{equation}
where $\bm\omega_1$ was defined in \eqref{eq.symcurrent}. Integrating this expression over a Cauchy slice $\Sigma$, one gets the Hamilton equations corresponding to a $\zeta$ diffeomorphism
\begin{equation}
\label{eq.hamiltoneq}
    \delta Q[\zeta]=-\Omega(\delta_\zeta\phi,\delta\phi)+\Omega(\delta\phi,\delta_\zeta\phi).
\end{equation}

Before concluding this section, we want to stress that the result we just reviewed, even if very important, will prove to be too restrictive for our needs, and it has to be generalized in different directions. In fact, Equation~\eqref{eq.hamiltoneq}
holds as long as we stay within $\mathcal P$, that is to say we consider perturbations that are tangent to the space of solutions of the field equations ($\mathbf{E}_0=0$) and satisfy the boundary conditions ($\mathbf{B}_1=0$).

One of the generalizations that we will need is straightforward. Let us consider a variation that is only tangent to $\Pi$ and not $\mathcal C$\,---\,and thus moves from a certain configuration space, defined by a boundary value of $\phi$, to a different one, defined by a slightly different value $\phi+\delta\phi$. In this case, the $\mathbf{B}_1$ term in $\mathbf{J}_2[\zeta]$ contributes a boundary term
\begin{equation}
\label{eq.localhamiltoneqviolatingbc}
    \delta(\mathbf{J}_1[\zeta]-\dd\mathbf{J}_2[\zeta])|_{\Pi}=-\bm{\omega}_1(\delta_\zeta \phi,\delta g)+\bm{\omega}_1(\delta \phi,\delta_\zeta \phi)+\dd(\bar\zeta\cdot \mathbf{B}_1).
\end{equation}

\section{Review of Lanczos-Lovelock theories}
\label{s.llt}

In this section, we present a short introduction to LL theories to make the paper self-contained. Our goal is to apply the formalism of covariant phase space to study black holes and their thermodynamics in this class of theories.

Lanczos-Lovelock theories were born as a generalization of GR,
when Lovelock set out to find the most general set of Lagrangians that produce second order field equations for the metric tensor \cite{Lanczos:1938sf, Lovelock:1971yv}. The resulting structure is that of a sum of terms with arbitrary coefficients.
Most results hold independently for each term, irrespective of its coefficient, so we will not include the sum symbol unless the validity of the result depend on the contribution of all terms.
Interest in these models sparked when they were found to describe low energy limit of supersymmetric string theory~\cite{Boulware:1985wk}.

To present LL theories, we separate the bulk Lagrangian as $\mathbf{L}_0=L_0\bm\epsilon_{\mathcal M}$, where $L_0$ is a scalar and $\bm\epsilon_{\mathcal M}$ is the volume form of spacetime. One has
\begin{equation}
\label{eq.bulklagrangian}
    L_0=\sum_k c_k  L_0^{(k)},\quad \text{with}\quad   L_0^{(k)}=\frac{1}{2^k}\delta^{\mu_1 \nu_1\dots \mu_k \nu_k}_{\rho_1 \sigma_1\dots \rho_k \sigma_k}\prod_{i=1}^k R^{\rho_i \sigma_i}_{\mu_i \nu_i},
\end{equation}
where $\delta^{\mu_1 \nu_1\dots \mu_k \nu_k}_{\rho_1 \sigma_1\dots \rho_k \sigma_k}$ is the generalized Kronecker delta, antisymmetric in all upper and lower indices, and $c_k$ are arbitrary constant coefficients. This Lagrangian has been proven to be the most general constructed out of the Riemann tensor (but not its derivatives) that yields field equations for the metric which are second-order. The index $k$ can take an arbitrary set of values between $0$ and $\floor{d/2}$. Beyond that value, the antisymmetry of the the generalized Kronecker delta would make the contribution vanish anyway. See~\cite{Padmanabhan:2013xyr} for an extensive review.

It turns out very useful to introduce the Lovelock tensors
\begin{equation}
    {P^{(k)}}^{\mu\nu}_{\rho\sigma}\equiv\fdv{  L_0^{(k)}}{R^{\rho\sigma}_{\mu\nu}}\eval_{g_{\alpha\beta}}=\frac{k}{2^k}\delta^{\mu\nu \mu_2 \nu_2\dots \mu_k \nu_k}_{\rho\sigma\rho_2 \sigma_2\dots \rho_k \sigma_k}\prod_{i=2}^k R^{\rho_i \sigma_i}_{\mu_i \nu_i},
\end{equation}
which have the same symmetries as the Riemann tensor and are covariantly conserved $\nabla_\mu {P^{(k)}}^{\mu\nu }_{\rho\sigma}=0$ (this property follows from the generalized delta antisymmetry and the differential Bianchi identity $\nabla_{[\lambda} R^{\rho\sigma}_{\mu\nu]}=0$). Since each Lagrangian term is a homogenous polynomial in the Riemann tensor, one can conveniently express them as
\begin{equation}
     L_0^{(0)}=1,\qquad L_0^{(k>0)}=\frac{1}{k}{P^{(k)}}^{\mu\nu}_{\rho\sigma} R^{\rho\sigma}_{\mu\nu}.
\end{equation}

The field equations can be written in terms of the generalized Einstein tensor\footnote{Notice that each generalized Einstein tensor admits the compact expression
\begin{equation}
    {\mathcal G^{(k)}}^\alpha_\beta=-\frac{1}{2^{k+1}}\delta^{\alpha\mu_1\nu_1\dots \mu_k\nu_k}_{\beta\rho_1\sigma_1\dots\rho_k\sigma_k}\prod_{i=1}^k R^{\rho_i \sigma_i}_{\mu_i \nu_i}.
\end{equation}}
\begin{equation}
\label{eq.gtensor}
    \mathcal G_{\alpha\beta}=\sum_k c_k \mathcal G^{(k)}_{\alpha\beta},\quad \text{with}\quad {\mathcal G^{(k)}}^\alpha_\beta\equiv{P^{(k)}}_{\beta \sigma}^{\mu\nu} R_{\mu\nu}^{\alpha \sigma}-\frac{1}{2}\delta^\alpha_\beta  L_0^{(k)}.
\end{equation}
Just like their usual counterpart, each Einstein tensor $\mathcal G^{(k)}_{\alpha\beta}$ is symmetric and divergenceless. When matter is present, the field equations are $\mathcal G_{\alpha\beta}= (1/2)T_{\alpha\beta}$, where $T_{\alpha\beta}$ is matter stress-energy tensor. In this work, we will be only concerned with vacuum solutions $\mathcal G_{\alpha\beta}= 0$.

As we previously mentioned, the distinctive feature of LL theories is that the field equations are second order in the metric tensor, despite the action functional might contain high-order monomials in the curvature. This property stems from the divergenceless condition on the Lovelock tensors.
Since the field equations are second order, we expect to be able to find a one-to-one correspondence between the initial values of the field and its derivative on the sub-manifold $\Sigma_i$ and the solutions to the field equations~\cite{Choquet-Bruhat:1988jdt,Deruelle:2003ck}. A well-posed initial value formalism ensures that the Hamiltonian constructed in Section~\ref{ss.diffinv} is sufficiently smooth to generate a good flow.

If the variational principle is to be satisfied, we have to remove boundary terms which are not eliminated by the boundary conditions. In this work, we choose to impose Dirichlet boundary conditions on $\Gamma$ 
\begin{equation}
    q\equiv j^*_\Gamma g\quad \text{fixed}.
\end{equation}
Therefore, the spurious terms are those that depend on derivatives of $g_{\alpha\beta}$ which are normal to the boundary. These terms are contained in the bulk symplectic potential, that we write as the contraction of its dual vector $\Theta_1$ with the spacetime volume form.\footnote{In components
\begin{equation}
    \mathbf{\Theta_1}=\frac{1}{(d-1)!}\Theta_1^\mu\,\epsilon_{\mu\mu_2\dots \mu_d}\dd x^{\mu_2}\wedge\dots\wedge\dd x^{\mu_d}.
\end{equation}} The bulk symplectic potential for the LL Lagrangian \eqref{eq.bulklagrangian} is
\begin{equation}
\label{eq.bulksympot}
    \Theta_1=\sum_k c_k  \Theta_1^{(k)},\quad \text{with}\quad   {\Theta_1^{(k)}}^\mu={P^{(k)}}^{\mu\nu}_{\rho\sigma}\pqty{g^{\rho \lambda}\nabla^\sigma-g^{\sigma \lambda}\nabla^\rho}\delta g_{\lambda \nu}.
\end{equation}
As in standard GR with Dirichlet boundary conditions, we can get rid of these terms adding a suitable boundary Lagrangian $\mathbf{L}_1=L_1\bm\epsilon_\Gamma$ (where $\bm\epsilon_\Gamma$ is the boundary volume element), which, in that case, is known as Gibbons-Hawking-York term.
To preserve covariance, this Lagrangian has to be constructed only from the boundary intrinsic and extrinsic geometry~\cite{Wald:1984rg, Carroll:2004st}. Indeed, 
\begin{equation}
\label{eq.boundlag}
    L_1=\sum_k c_k  L_1^{(k)},\quad \text{with}\quad  L_1^{(k)}=\frac{2k}{2^{k-1}}\int^1_0\dd s\,\delta^{i a_2 b_2\dots a_{k} b_{k}}_{j c_2 d_2\dots c_{k}d_{k}} K^j_i \prod_{i=2}^{k}\pqty{R^{c_i d_i}_{a_i b_i}-2s^2 K^{c_i}_{a_i} K^{d_i}_{b_i}},
\end{equation}
where $K_{ab}$
is the extrinsic curvature of $\Gamma$ and $R^{ab}_{cd}$ is the Riemann tensor constructed out of the induced metric $q\equiv j^*_\Gamma g$ \cite{Miskovic:2007mg, Padmanabhan:2013xyr}. Each term in the product is the curvature of the so-called \textit{interpolating connection}, which, as the name says, interpolates between the curvature of $\Gamma$ at $s=0$, and the full spacetime curvature projected on $\Gamma$ at $s=1$ (cfr. with Section $6.1$ of Ref.~\cite{EGUCHI1980213}). 

Indeed, in Appendix \ref{app.blct} we shall show that
\begin{equation}
\label{eq.boundlllagvar}
    \delta \mathbf{L}_1=\frac{1}{2}t_{ab}\delta q^{ab}\bm{\epsilon}_\Gamma-j^*_\Gamma\bm\Theta_1+\dd\bm{\Theta}_2
\end{equation}
where $t_{ab}$ generalizes the well-known Brown-York tensor~\cite{Brown:1992br}
\begin{equation}
    t_{ab}=\sum_k c_k  t_{ab}^{(k)},\quad \text{with}\quad t^{(k)}_{ab}=4{\pi^{(k)}}^c_{(a} K_{b)c}
    -q_{ab} L^{(k)}_1-4(k-1)\int_0^1\dd s\, {p^{(k)}}^{ice}_{jd{(a}}(s) K^j_i R_{b)}{}^d{}_{ce},
\end{equation}
and $\bm\Theta_2$ is the corner symplectic potential, obtained from the contraction of its dual vector $\Theta_2$ with the volume form of the boundary.\footnote{In components
\begin{equation}
    \bm\Theta_2=\frac{1}{(d-2)!}\Theta_2^a\, \epsilon_{a a_2\dots a_{d-1}}\dd y^{a_2}\wedge\dots\wedge\dd y^{a_{d-1}}.
\end{equation}} Its expression is
\begin{equation}
    \Theta_2=\sum_k c_k  \Theta_2^{(k)},\quad \text{with}\quad   {\Theta_2^{(k)}}^a=-2{\pi^{(k)}}^{ab}\!\not\!\delta c_b+4(k-1)\int_0^1\dd s\, {p^{(k)}}^{iab}_{jcd}(s) K^j_i q^{ce}q^{df}D_e\delta q_{bf}.
\end{equation}
which, to the best of our knowledge, has not been presented in this novel and concise form before.\footnote{After the completion of this paper, we became aware that an expression for the corner symplectic potential was also obtained in Refs.~\cite{Chakraborty:2017zep, Cano:2018ckq}. Our potential agree with that of Ref.~\cite{Cano:2018ckq} by direct inspection, while the one found in Ref.~\cite{Chakraborty:2017zep} is more difficult to compare. In fact, the authors of ~\cite{Chakraborty:2017zep} express the result in terms of the sum one gets by expanding the product inside $p^{(k)}(s)$ and integrating each resulting monomial over $s$. It would be interesting to perform these computations and check if we get compatible results.}
In these formulas, to simplify the notation, we introduced the tensor
\begin{equation}
{p^{(k)}}^{iab}_{jcd}(s)=
\begin{cases}
    0\quad& \text{for}\quad k=0,1,\\
    \delta^{iab}_{jcd}\quad &\text{for}\quad k=2,\\
    \frac{k}{2^{k-1}}\delta^{iaba_3b_3\dots a_{k}b_{k}}_{jcdc_3d_3\dots c_{k}d_{k}}\prod_{i=3}^{k} \pqty{R^{c_i d_i}_{a_i b_i}-2s^2  K^{c_i}_{a_i}  K^{d_i}_{b_i}}\quad &\text{for}\quad k>2,\\
\end{cases}
\end{equation}
and the tensors ${\pi^{(k)}}^a_b$, $\!\not\!\delta c_a$, which are respectively the pull-back to $\Gamma$ of
\begin{equation}
    {\pi^{(k)}}^\alpha_\beta\equiv n_\mu n^\nu {P^{(k)}}^{\alpha\mu}_{\beta\nu},\qquad \!\not\!\delta c_\alpha\equiv \delta g_{\mu\nu}n^\nu q^\mu_\alpha.
\end{equation}
Notice that $t_{ab}\delta q^{ab}$ vanishes as required if we impose Dirichlet boundary conditions.

Let us take a moment of pause to see that these results indeed recover what we known in the familiar case of GR~\cite{Harlow:2019yfa}
\begin{align}
    &t^{(1)}_{ab}=2(K_{ab}-K q_{ab}),\qquad {\Theta_2^{(1)}}^a=-\!\not\!\delta c^a.
\end{align}

\subsection{Pure and generic theories}
\label{ss.pureandimpure}

It is easy to see from \eqref{eq.gtensor} that the trace of the vacuum field equations implies
\begin{equation}
\label{eq.fieldeqtrace}
    \sum_k c_k (d-2k) L_0^{(k)}= 0
\end{equation}
Unlike the case of GR, one cannot conclude from this that the bulk Lagrangian vanishes itself (unless each term vanishes independently). The only case in which one could is if only one of the constant was different from $0$. This motivates the following distinction: a LL theory is ``pure'' if there is only one non-vanishing coefficient, otherwise it is ``generic''.

Pure theories have a lot more in common with GR, especially in their geometrical description. Indeed, for any $k$ term, one can build a generalization of the Riemann tensor which satisfies the differential Bianchi identity~\cite{Dadhich:2008df, Kastor:2012se}. It has been proven that, for pure theories, vacuum solutions are flat with respect to this generalized Riemann tensor in the odd-dimension $2k+1$~\cite{Dadhich:2012cv}.

Another property of pure theories is that their vacuum, that is the maximally symmetric solution, is just flat spacetime. On the other hand, generic LL admit cosmological vacuum solutions, even in the absence of the actual cosmological constant term ($L_0^0$). To see this, we can rewrite the Einstein tensor as
\begin{equation}
    \mathcal G^\alpha_\beta=-\sum_{k=0}^m \frac{c_k}{2^{k+1}}\delta^{\alpha\mu_1\nu_1\dots \mu_k\nu_k}_{\beta\rho_1\sigma_1\dots\rho_k\sigma_k}\prod_{i=1}^k R^{\rho_i \sigma_i}_{\mu_i \nu_i}=-\frac{c_m}{2^{m+1}}\delta^{\alpha\mu_1\nu_1\dots \mu_m\nu_m}_{\beta\rho_1\sigma_1\dots\rho_m\sigma_m}\prod_{i=1}^m (R^{\rho_i \sigma_i}_{\mu_i \nu_i}-\alpha_i \delta^{\rho_i \sigma_i}_{\mu_i \nu_i}),
\end{equation}
for appropriate coefficients $\alpha_1,\dots,\alpha_m$ which are determined in terms of combinations of the original coefficients $c_0,\dots, c_m$. Whereas these combinations are usually complex, some of them might be real. In those cases, the theory admits up to $m$ constant curvature vacua
\begin{equation}
    R^{\rho \sigma}_{\mu \nu}=\alpha_{I} \delta^{\rho \sigma}_{\mu \nu},\quad \forall\alpha_I\in\mathds R.
\end{equation}

\subsection{Lanczos-Lovelock diffeomorphism charges}

Generally covariant theories have a large degree of redundancy. Using Noether's second theorem, one expects the equations of motion to not be completely independent. Indeed, in this case, we have a generalization of the usual Bianchi identities $\nabla_\mu \mathcal G^{\mu\nu}=0$. For these theories, the Noether current admits an improvement which is closed also off-shell~\cite{Wald:1990mme}. Indeed, from \eqref{eq.bulkcurrentconservation} and using the Bianchi identity, we get
\begin{equation}
\label{eq.bulkcurrentconservationLL}
    \dd\mathbf{J}_1[\zeta]=\mathcal G^{\alpha\beta}\delta_\zeta g_{\alpha\beta}\bm\epsilon_{\mathcal M}=2\nabla_\alpha (\mathcal G^\alpha_\beta \zeta^\beta)\bm\epsilon_{\mathcal M},
\end{equation}
that we can rewrite as (minus) the external derivative of the form $\mathbf{K}_1\equiv K_1\cdot\bm\epsilon_{\mathcal M}$ with $K_1^\mu=-2\mathcal G^\mu_\nu \zeta^\nu$, called the constraint current. Therefore, as long as spacetime satisfies the hyphotesis of Poincaré lemma (basically simply connectedness), one can can construct a potential such that $\mathbf J_1+\mathbf{K}_1= \dd\mathbf q_2$. We can write this potential in terms of the antisymmetric tensor
\begin{equation}
    q_2^{\mu\nu}[\zeta]=-\sum_k c_k {P^{(k)}}^{\mu\nu}_{\rho\sigma}(\nabla^\rho\zeta^\sigma-\nabla^\sigma\zeta^\rho).
\end{equation}
An immediate consequence is that, for generally covariant theories, the total Noether charge \eqref{eq:NoetherCh}, associated to diffeomorphisms is a pure boundary term
\begin{equation}
\label{eq.holographiccharge}
    Q[\zeta]=\int_{\partial \Sigma} \pqty{\mathbf{q}_2[\zeta]-\mathbf{J}_2[\zeta]}.
\end{equation}

Reproducing the argument of~\cite{Harlow:2019yfa}, we now derive an alternative formula for the Noether charge which does not depend on normal derivatives of the vector field $\zeta$.
As we described in Sec.~\ref{ss.diffinv}, the presence of the boundary $\Gamma$ restricts the group of allowed diffeomorphisms to those satisfying Eq.~\eqref{eq.conditions}. In particular, the second of those conditions tells us that $\zeta$ must approach a Killing symmetry of $\Gamma$
\begin{equation}
\label{eq.conditionb}
    \delta_\zeta (j^*_\Gamma g)=\delta_{\bar\zeta} q=0.
\end{equation}
Let us go back to \eqref{eq.chargeconservation} and express the boundary metric variation as the covariant (with respect to $q$) derivative of $\bar\zeta$. Since the generalized Brown-York is symmetric and conserved in the absence of matter fluxes (this follows from the invariance of the action under large diffeomorphisms), we can integrate by part to express the last integral as the difference between two corner terms
\begin{equation}
    \Delta Q_{f/i}[\zeta]=-\int_{\Gamma} \frac{1}{2}t_{ab}\delta_{\bar\zeta} q^{ab}\bm{\epsilon}_\Gamma=-\int_{\partial\Sigma_f} u_a {t^a}_b\bar\zeta^b \bm{\epsilon}_{\partial\Sigma}+\int_{\partial\Sigma_i} u_a  {t^a}_b\bar\zeta^b \bm{\epsilon}_{\partial\Sigma}.
\end{equation}
In the previous expression, we used $\bm\epsilon_\Gamma=\bm u\wedge \bm\epsilon_{\partial\Sigma}$, where $u$ is the unit normal $1$-form to $\partial\Sigma$.
This means that the charge can be expressed (up to a constant) as the integral of the Brown-York current
\begin{equation}
\label{eq.bycharge}
    Q_{BY}[\zeta]=-\int_{\partial\Sigma} u_a {t^a}_b\bar\zeta^b \bm{\epsilon}_{\partial\Sigma}.
\end{equation}
From this result, one can easily see that small diffeomorphisms\,---\,those that vanish on $\Gamma$\,---\,yield no conserved charge. This is associated with small diffeomorphisms being redundancies of the theory rather than actual symmetries. One can easily see from \eqref{eq.bycharge} that such diffeomorphisms yield no conserved charge.

\section{Black holes thermodynamics}
\label{s.bhs}

Finding black hole solutions in Lanczos-Lovelock proved to be quite an hard task. The first exact solution was found in \cite{Boulware:1985wk}, motivated by low energy supersymmetric string theory considerations. After that, many other solutions have been found (see Refs~\cite{Myers:1998gt} and \cite{Garraffo:2008hu} for a nice review). 
Moreover, results have been obtained in the study of Birkhoff's theorem in LL theories, at least for the particular class known as Einstein-Gauss-Bonnet (cfr. Section~\ref{ss.bh5d}) \cite{Deser:2003up, Deser:2005gr, Zegers:2005vx}.

However, important milestones of our understanding of black holes do not survive in more than $4$ dimensions, where LL become interesting. It is known that Hawking topology theorem \cite{Hawking:1971vc} does not hold in higher dimensions \cite{Galloway:2005mf}. The rigidity theorem \cite{Hawking_Ellis_1973} has been generalized to prove the existence of one rotational isometry for stationary black holes \cite{Hollands:2006rj}, but it is unclear whether the result can be made stronger. See \cite{Emparan:2008eg, Hollands:2005wt} for a review.

For the purpose of our paper, we will initially consider black hole geometries which are stationary, possess some rotational isometries and are non-extremal. 
Later on, when deriving the Smarr formula, we will restrict our attention to static, spherically symmetric solutions.

\subsection{First law of thermodynamics}
\label{ss.firstlaw}

In deriving the first law, we limit ourselves to stationary perturbations of stationary (non-extremal) black holes, which means that we consider solutions of the field equations which admit a Killing vector field $\xi$ (with non-zero surface gravity) and restrict to perturbations such that $\delta\xi=0$. We shall further assume that these solutions have at least one Killing horizon, where $\xi$ becomes null, and assume that $\xi$ is timelike outside such horizon.\footnote{If there are multiple horizons, we always consider the outer one beyond which $\xi$ becomes timelike.} As we previously mentioned, we further assume the Killing vector field can be decomposed as $\xi=t+ \Omega^i_H \varphi^i$ (the sum over $i$ is implicit), where $t$ is the Killing vector field generating boundary time translations and $\varphi^i$ is a collection of Killing vector fields generating rotations, and $\Omega_H^i$ is a collection of constants that one can identify as the horizon angular velocities corresponding to the $i$-th rotation.

Given the form of \eqref{eq.bycharge}, we are naturally led to define canonical charges associated to the diffeomorphisms $t$ and $\varphi^i$ which have the interpretation of canonical energy and canonical angular momenta
\begin{gather}
    \label{eq.defcanonicalenergy}
    E\equiv -\int_{\mathcal S} u_a {t^a}_b t^b \bm{\epsilon}_{\mathcal S},\\
    \label{eq.defcanonicalspin}
    J^i\equiv \int_{\mathcal S} u_a {t^a}_b(\varphi^i)^{\,b} \bm{\epsilon}_{\mathcal S}.
\end{gather}

We are now in the position to formulate the first law of black hole thermodynamics within the covariant phase space formalism.
To derive a the law for stationary black holes, we start from the local relation \eqref{eq.localhamiltoneq} and notice that the r.h.s.~vanishes trivially since $\delta_\xi g=\mathcal L_\xi g=0$
\begin{equation}
    \delta(\mathbf{J}_1[\xi]-\dd\mathbf{J}_2[\xi])=-\bm{\omega}_1(\delta_\xi g,\delta g)+\bm{\omega}_1(\delta g,\delta_\xi g)=0.
\end{equation}
Next, we integrate this relation over a slice $\Sigma$, but instead of taking $\Sigma$ to be a Cauchy hypersurface as we did to derive \eqref{eq.hamiltoneq}, we take it to be a generic slice that extends from the horizon bifurcation surface $\mathcal B$ (that we assumed to exist) to $\mathcal S$.
The integrals at $\mathcal S\subset\Gamma$ reproduce the variations of the canonical charges $E$ and $J^i$, while the integral at $\mathcal B$ can be written as the variation of~\footnote{Notice that $\mathbf J_2[\xi]$ has no support on $\mathcal B$.}
\begin{equation}
\label{eq.waldentropy}
    S\equiv 2\pi \int_{\mathcal B}\mathbf{q}_2[\xi/\kappa],
\end{equation}
which agrees with Wald's prescription for the entropy~\cite{Wald:1993nt}. Overall, we get
\begin{equation}
\label{eq.firstlaw}
    \delta E=\Omega^i_H \delta J^i+\frac{\kappa}{2\pi}\delta S,
\end{equation}
where $\kappa$ is the surface gravity of $\xi$. This relation is usually presented with the Clausius term $T_H\delta S$, where $T_H$ is the Hawking temperature. However, in our treatment, there are two main differences: (i) the Hawking temperature is usually given as a property of the black hole without boundary, so the actual temperature $T$ differs from the Hawking one $T_H$ because of finite-size effects, (ii) since we do not normalize $\xi$ at the boundary, we have to use the full Tolman relation $\norm{\xi} T=\frac{\kappa}{2\pi}$.

It is worth noticing here that the above defined entropy for LL theories does not coincide with the Bekenstein-Hawking one, for it is not simply $1/4$ of the surface area of the horizon, but also includes a sum of (integrated) intrinsic curvature invariants of a cross section of the horizon~\cite{Jacobson:1993xs}.
As a matter of fact, Wald entropy is in this case just the action of a LL theory defined on the bifurcation surface, with coefficients that can be obtained from the original ones
\begin{equation}
\label{eq.noethercharge}
    S=-4\pi\int_{\mathcal B}\sum_{k} k c_k \mathbf{L}_0^{(k-1)},
\end{equation}
as one can show using the relation $\nabla_{[\mu} \xi_{\nu]}=-2\kappa u_{[\mu} n_{\nu]}$. Since we are dealing with stationary horizons, the several ambiguities~\cite{Jacobson:1993vj} that affect the entropy definition do not alter Wald's result.

As we saw, we have been able to define canonical charge that satisfy the first law without the necessity to postulate any ``integrability condition". In the early approaches, boundary quantities were not taken into account, so that the Noether current and the symplectic potential were respectively given by $\mathbf{J}_1$ and $\bm\Theta_1$ only. Accordingly, people expected the flow of the symplectic current $\delta\bm\Theta_1$ (the first two terms in \eqref{eq.bulkcurrentvariation}) to be integrable by itself. This requires the existence of a codimension $2$ form $\mathbf{B}_2(\xi)$\,---\,not to confused with $\mathbf{B}_1$ in \eqref{eq.boundlagvar}\,---\,such that $\delta \mathbf B_2(\xi)=\xi\cdot \bm\Theta_1$. The solution given by a careful application of the covariant phase space algorithm is that the contraction with the diffeomorphism-induced vector field of the true symplectic current $\bm\omega_1$ is always integrable thanks to the contribution of the boundary Noether current. In fact, it is $\mathbf{J}_2$ the form that unambiguously plays the role of $\mathbf{B}_2(\xi)$.

\subsection{Scale invariance breaking in Lanczos-Lovelock theories}

In this section, we review the problem of extending the Smarr formula to LL theories beside GR, and then provide a simple and clean derivation of it from the first law of thermodynamics using the prescription of covariant phase space formalism to construct the needed terms.

There are two ways to derive the Smarr formula in GR: one is by integrating the first law of thermodynamics for the variation induced on macroscopic quantities by an overall change of length scale, the other is by means of the famous Komar integral relation~\cite{Komar}, which relies on the existence of a non-trivial quantity that is divergenceless on-shell and thus yield an equality between quantities evaluated on different surfaces ($\mathcal S$ and $\mathcal B$ to be precise).

As mentioned in Section~\ref{ss.Introduction}, the reason why it took a long time to extend the Smarr formula to generic LL theories is the breaking of scale invariance. For vacuum GR without cosmological constant, one can set Newton constant to be unity from the beginning and get a theory with no dimensional couplings. As a result, the field equations are invariant under a constant Weyl scaling of the metric $\delta_W g_{\alpha\beta}=2g_{\alpha\beta}$
\begin{equation}
    \mathcal G_{\alpha\beta}=0 \implies \delta_W \mathcal G_{\alpha\beta}=0.
\end{equation}
In this case, on-shell quantities like energy and entropy can only depend on dynamical scales and thus the way they change under a constant Weyl scaling is determined by their length dimension. The Smarr formula is then a consequence of Euler's theorem for homogeneous functions.
These considerations hold true for any pure LL theory (see Section~\ref{ss.pureandimpure} for the definition), but we just focused on GR because it is by far the most studied pure LL.

On the other hand, as first noticed in~\cite{Gibbons:2004ai} and then elaborated in~\cite{Kastor:2008xb}, generic LL theories\,---\,among which GR with a cosmological constant\,---\,break scale invariance. Since the $k$-th generalized Einstein tensor $\mathcal G^{(k)}_{\alpha\beta}$ scale as $2(1-k)$ under Weyl transformation, the total generalized Einstein tensor does not scale homogeneously, thus precluding the aforementioned route for deriving the Smarr formula.

\subsubsection{Recovering scale invariance through extended thermodynamics}
\label{ss.recoveringsi}

Originally, the Smarr formula for LL was derived as a natural generalization of the case of GR with a cosmological constant~\cite{Kastor:2008xb} using the Komar relation. However, the physical insight that comes with the scaling argument is much more profound and we prefer to start with it.
In this section, we combine the prescription of the covariant phase space formalism and the ideas of the extended thermodynamic framework~\cite{Kastor:2009wy, Kastor:2010gq} to obtain the Smarr formula from the first law of black hole thermodynamics.

From the formal point of view, we cannot expect \eqref{eq.localhamiltoneq} to work for a constant Weyl scaling because $\delta_W$ is not tangent $\Pi$. The solution to this problem is to consider an extended framework in which variations are allowed to act also on couplings. To distinguish these new variations from the original ones, which only act on dynamical field, we use the symbol $\dbar$. Under a length scale change, we define the action of $\dbar_W$ on the Lovelock couplings to be defined by their length dimension $\dbar_W c_k\equiv(2k-d)c_k$. In this way, the products $c_k \mathcal G^{(k)}_{\alpha\beta}$ transform homogeneously under $\dbar_W$ irrespective of $k$, and the space of solutions is left invariant by the extended variation.

Let us see how the first law \eqref{eq.firstlaw} changes in the extended thermodynamic framework. For the sake of presentation, we first restrict to variations that are tangent to the configuration space $\mathcal C$. Therefore, we can work with Eq. \eqref{eq.localhamiltoneq}. The advantage of our derivation is that it can be done in two lines.
First, we factor out the $c_1$ coefficient and use the Leibniz rule to rewrite the l.h.s.~of \eqref{eq.localhamiltoneq} as
\begin{equation}
\label{eq.passaggio1}
    \frac{1}{c_1}\delta (\mathbf{J}_1-\dd\mathbf{J}_2)=\sum_k \frac{c_k}{c_1} \delta (\mathbf{J}^k_1-\dd\mathbf{J}^k_2)=\sum_k \dbar \pqty{\frac{c_k}{c_1} (\mathbf{J}^k_1-\dd\mathbf{J}^k_2)}-\sum_k (\mathbf{J}^k_1-\dd\mathbf{J}^k_2)\dbar\pqty{\frac{c_k}{c_1}}
\end{equation}
which must be equated to the r.h.s.~of \eqref{eq.localhamiltoneq} which is $0$.
Then, as we did in the non-extended case, we integrate this relation on a slice that connects the bifurcation surface $\mathcal B$ to the boundary section $\mathcal S=\Sigma\cap\Gamma$ to obtain
\begin{equation}
\label{eq.extendedfirstlaw}
    \dbar (c_1^{-1}E)-\Omega^i_H\dbar(c_1^{-1}J^i)-\frac{\kappa}{2\pi} \dbar(c_1^{-1} S)-\sum_k \Psi_k\, \dbar (c_1^{-1}c_k)= 0
\end{equation}
where we denoted with $\Psi_k$ the $k$-th component of the total Noether charge associated to the Killing isometry~\footnote{We used again that $\mathbf{J}^{(k)}_2[\xi]$ is absent on $\mathcal B$.}
\begin{equation}
\label{eq.thermopotentials}
    \Psi_k\equiv \int_\Sigma \mathbf{J}^k_1[\xi]-\int_{\mathcal S}\mathbf{J}^k_2[\xi].
\end{equation}
With this quick derivation, we were able to obtain a prescription for the thermodynamical potentials conjugated to the (relative) Lovelock couplings that is simple and unambiguous.

However, in order to derive the Smarr formula, we have to generalize further and consider the extended first law of thermodynamics for variations that violate the boundary conditions given that $\delta_W q^{ab}=-2q^{ab}\ne 0$. 
Let us then repeat the previous argument starting from \eqref{eq.localhamiltoneqviolatingbc} rather than \eqref{eq.localhamiltoneq}: the l.h.s. that we find will be the same as \eqref{eq.passaggio1}; but the r.h.s., even if $\delta_\xi g=0$, will not be $0$ because of the boundary term. Using the fact that $\bm B_1$ has no support on $\mathcal B$, that boundary term yields
\begin{equation}
\label{eq.passaggio2}
    \int_\Sigma\dd \pqty{\bar\xi\cdot\mathbf B_1}\overset{\eqref{eq.boundlllagvar}}{=}\frac{1}{2}\int_{\mathcal S}t_{ab}\delta q^{ab} (\bar\xi\cdot\bm\epsilon_\Gamma),
\end{equation}
up to the $c_1$ that we factored out.
We evaluate this term by performing a $1+(d-2)$ decomposition of the boundary metric. We choose the vector normal to $\mathcal S\subset\partial\Sigma$ to be proportional to the generator of time translations $t$ and introduce $\lambda\equiv\norm{t}=\sqrt{-t^2}$.
All the $\varphi^i$ vectors are tangent to $\mathcal S$ and thus drop from the contraction $\bar\xi\cdot\bm\epsilon_\Gamma$. Splitting $q^{ab}=-u^a u^b+\gamma^{ab}$, with $\gamma$ the projector on $\mathcal S$, we study each term separately.
\begin{itemize}
    \item The variation of the first part $u^a u^b$ can be rewritten as $2\, t^a u^b \delta \lambda^{-1}$ using the proportionality between $t$ and $u$, and assuming that the perturbations are such that $\delta t=\delta \varphi^i=0$ at the boundary.\footnote{This condition is independent from the stationary perturbations condition $\delta\xi=0$ that we imposed before.} Similarly, by linearity, we can replace $t\cdot \bm\epsilon_\Gamma$ with $\lambda u\cdot \bm\epsilon_\Gamma=-\lambda\bm\epsilon_{\mathcal S}$ 
\begin{equation}
    \frac{1}{2}\int_\mathcal S t_{ab}\delta(-u^a u^b)(t\cdot\bm\epsilon_\Gamma)=\int_\mathcal S (-t_{ab} t^a u^b \delta\lambda^{-1}) (-\lambda \bm\epsilon_{\mathcal S})=-\int_\mathcal S (\lambda^{-1}\delta\lambda) t_{ab} t^a u^b \bm\epsilon_\mathcal S.
\end{equation}
We recognize that, if it were not for the term in brackets (which merely account for a variation of the boundary observer's proper time), the last integral would equate the canonical energy as defined in \eqref{eq.energy}. So, with some license, we indicate this integral as $(\lambda^{-1}\delta\lambda)E$, with the caveat that, as of now, this is not a product.

\item The variation of the second part is not as simple but we can think of it as a work term related to changes of the boundary surface. We thus write it as $-\tilde\Sigma\delta A$ with the same license as before.
To justify this notation, let us consider the case that $t_{ab}$ has a surface tension part $-\sigma\gamma_{ab}$; then, it easily follows that
\begin{equation}
\label{eq.surfacetensionterm}
    \frac{1}{2}\int_\mathcal S t_{ab}\delta\gamma^{ab}(t\cdot\bm\epsilon_\Gamma)=-\frac{1}{2}\int_\mathcal S \sigma\,\gamma_{ab}\delta \gamma^{ab} (-\lambda \bm\epsilon_{\mathcal S})=-\int_S \lambda\sigma\, \delta \bm\epsilon_{\mathcal S},
\end{equation}
so, if the surface tension $\lambda\sigma\equiv\tilde\Sigma$ were constant on $\mathcal S$, we would exactly have the product $\tilde\Sigma\delta A$, with $A$ the surface area of $\mathcal S$.
\end{itemize}

Summing these two terms and equating them to the l.h.s. of \eqref{eq.extendedfirstlaw}, we can write an extended version of the first law which generalizes \eqref{eq.localhamiltoneqviolatingbc} to variations that are neither tangent to $\mathcal C$ nor to $\Pi$
\begin{equation}
\label{eq.presmarr}
     \dbar (c_1^{-1}E)-\Omega^i_H \dbar(c_1^{-1}J^i)-\frac{\kappa}{2\pi} \dbar(c_1^{-1} S)-\sum_k \Psi_k\, \dbar (c_1^{-1}c_k)= (\lambda^{-1} \delta \lambda) c_1^{-1}E-c^{-1}_1\tilde\Sigma\delta A.
\end{equation}

Let us now specialize this equation for a change in length scale. The length dimension of canonical charges and entropy (divided by $c_1$) is $d-2$, while the scaling of the ratio $c_k/c_1$ is $2(k-1)$; the clock variation $\lambda^{-1}\delta_W\lambda$ gives just $1$, while the area scales unsurprisingly as $d-2$. Finally, reinstating $c_1$, we arrive at the main result of this paper
\begin{equation}
\label{eq.finitesizesmarr}
     (d-3)E-(d-2)\Omega^i_H J^i-(d-2)\frac{\kappa}{2\pi} S-2\sum_k (k-1)c_k \Psi_k= -(d-2)\tilde\Sigma A,
\end{equation}
where we wrote $\tilde\Sigma A$ instead of the integral
\begin{equation}
    -\frac{\lambda}{d-2}\int_\mathcal S  t_{ab}\gamma^{ab}\bm\epsilon_{\mathcal S}.
\end{equation}

While interesting, this equation is not exactly a Smarr formula, as it involves quantities which are not intrinsic to the black hole, but are related to the position and motion of the boundary observer. In the next Section, after the discussion about Komar integrals, we will show how to get rid of observer's dependent effects and derive the Smarr formula.

\subsubsection{Komar integrals}

An alternative way to obtain the Smarr formula is to start from a Komar integral, which is an integral evaluated on the boundary of a spatial slice that vanishes by virtue of the field equations
\begin{equation}
\label{eq.komarrelation}
    \int_{\partial\Sigma} \mathbf{a}_2 = 0
\end{equation}
By choosing an hypersurface $\Sigma$ stretching from the horizon to the boundary, the previous equation would establish a relation between entropy and canonical charges.

Let us get back to \eqref{eq.bulkcurrentconservationLL} and notice that, for a Killing symmetry $\xi$, the term in the middle vanishes also for $\mathcal G^{\alpha\beta}\ne 0$, so even off-shell. This means that the constraint current associated to a Killing symmetry admits a potential $\mathbf{K}_1[\xi]=\dd \mathbf{c}_2[\xi]$. The existence of this potential that is closed on-shell is related to the possibility of integrating the equations of motion to get \eqref{eq.algfieldeq}. Indeed, the (constant) value of $\int_{\mathcal S}\mathbf{c}_2[\xi]$ is proportional to the ADM mass.
Given that we found a codimension $2$ form that is closed on-shell, it appears that we can realize \eqref{eq.komarrelation} straightforwardly by identifying $\mathbf{a}_2=\mathbf{c}_2[\xi]$. While this technically works, this does not produce any meaningful result, as the two integrals are trivially equal.
In order to find something meaningful, we first look at the subcase of GR, where the Komar integral is
\begin{equation}
    \int_{\partial\Sigma}\mathbf q_2[\xi]= \int_{\Sigma}\mathbf J_1[\xi]=-\int_{\Sigma}\xi\cdot \mathbf L_0= 0.
\end{equation}

As we see, this result is deeply tied to the fact that, for pure LL theories (cfr. Sec. \ref{ss.pureandimpure}), the on-shell bulk Lagrangian vanishes in the vacuum.\\
Since, for a generic theory, the best one can say about the on-shell bulk Lagrangian is \eqref{eq.fieldeqtrace}, one might think that there is no Komar relation.
However, in GR and other pure theories, the existence of a single $\mathbf{c}_2[\xi]$ makes its inclusion irrelevant to the Komar integral. On the other hand, in generic theories, one has various $\mathbf{c}^{(k)}_2[\xi]$ that can be combined with different coefficients in order to get something non-trivial.\footnote{Only the combination $\sum_k c_k \mathbf{c}_2^{(k)}[\xi]$ drops out of the Komar integral.}
Let us then express the Komar form as a linear combination of $\mathbf{q}^{(k)}_2[\xi]$ and $\mathbf{c}^{(k)}_2[\xi]$, and try to fix the coefficients appropriately
\begin{equation}
    \mathbf{a}_2=\sum_k c_k\pqty{\alpha_k \mathbf{q}^{(k)}_2[\xi]+\beta_k \mathbf{c}^{(k)}_2[\xi]}.
\end{equation}
Taking an external derivative, we get
\begin{equation}
    \dd\mathbf{a}_2=\sum_k c_k\pqty{\alpha_k \mathbf{J}^{(k)}_1+(\beta_k+\alpha_k) \mathbf{K}^{(k)}_1}=\sum_k c_k\pqty{-\alpha_k \xi\cdot\mathbf{L}^{(k)}_0+(\beta_k+\alpha_k) \mathbf{K}^{(k)}_1},
\end{equation}
Looking again at the trace of the field equations, we see that we can realize \eqref{eq.komarrelation} by setting $\alpha_k=-\beta_k\propto (d-2k)$. If we fix the the normalization to reproduce result of GR, we get the Komar relation
\begin{equation}
    \int_{\partial\Sigma}\sum_k c_k\frac{d-2k}{d-2}\pqty{\mathbf q_2^{(k)}[\xi]-\mathbf c_2^{(k)}[\xi]}= 0.
\end{equation}

Now, one would like to relate the quantities that appear in this integral to canonical charges. What we will do instead is to show the equivalence of this with \eqref{eq.finitesizesmarr}, from which the Smarr formula follows as we already showed. If we notice that $(d-2k)$ is the eigenvalue of both $\mathbf{q}_2^{(k)}[\xi]$ and $\mathbf{c}_2^{(k)}[\xi]$ under a change of length scale, we can rewrite the l.h.s. as
\begin{equation}
    \delta_W\int_{\partial\Sigma}\sum_k c_k\pqty{\mathbf q_2^{(k)}[\xi]-\mathbf c_2^{(k)}[\xi]}=\delta_W\int_{\partial\Sigma}\pqty{\mathbf q_2[\xi]-\mathbf c_2[\xi]}=\delta_W\int_\Sigma \mathbf{J}_1[\xi],
\end{equation}
or, in terms of the total Noether charge
\begin{equation}
    \delta_W Q[\xi]=\int_{\partial\Sigma}\delta_W\mathbf{J}_2[\xi].
\end{equation}
Using \eqref{eq.boundarycurrentvariation}, dropping the total derivative and using the antisymmetry of $\bm\Theta_1$, we can rewrite this relation as
\begin{equation}
    \delta_W Q[\xi]=-2\int_{\mathcal S}\bar\xi\cdot\mathbf{B}_1[\xi](g,g).
\end{equation}
The Noether charge variation under a constant Weyl scaling has been evaluated in the last section (cfr.\eqref{eq.passaggio1} and the l.h.s. of \eqref{eq.presmarr}). On the other hand, the boundary integral of $\bar\xi\cdot\mathbf{B}_1[\xi](g,g)$ is obtained as a particular case of \eqref{eq.passaggio2} when the variation is $\delta_W g^{ab}=-2g^{ab}$ (cfr. the r.h.s. of \eqref{eq.presmarr}). Therefore, the previous relation is equivalent to \eqref{eq.finitesizesmarr}.

\subsection{Schwarzschild-like solutions}

In this Section, we study Schwarzschild-like solutions, i.e. static and spherically symmetric metrics of the form
\begin{equation}
\label{eq.ansatz}
    \dd s^2=-e^{2\nu(r)} f(r)\dd t^2+\frac{\dd r^2}{f(r)}+r^2\dd\Omega_{d-2}^2,
\end{equation}
with $\dd\Omega_{d-2}^2$ the line element of a unit codimension $2$ sphere with area $\Omega_{d-2}=2\pi^{\frac{d-1}{2}}\Gamma\pqty{\frac{d-1}{2}}$ and curvature $(d-2)(d-3)$. Given the symmetries of the ansatz, the generalized Einstein tensor $\mathcal G^\alpha_\beta$ is diagonal and the spherical components are all equal. In solving the field equations, we notice that the combination $\mathcal G^t_t-\mathcal G^r_r$ can be made to vanish straightforwardly by setting $\nu$ to be constant. This constant is usually set to $0$ using the freedom to rescale the time coordinate, but we do not because we will use that freedom to fix the value of $g_{tt}$ at some particular radius as part of the boundary conditions. The remaining field equations are respectively first order and second order differential equations for $f$, but the latter is redundant because of the Bianchi identities. The former one can be integrated exactly, yielding the algebraic equation
\begin{equation}
\label{eq.algfieldeq}
    \sum_k \hat c_k \pqty{\frac{1-f}{r^2}}^k=\frac{M}{\Omega_{d-2}\, r^{d-1}},
\end{equation}
where we rescaled the couplings as $\hat c_k=c_k\,(d-2)!/(d-2k-1)!$ and fixed the integration constant in terms of the ADM mass. In the following, we will justify this identification.

To preserve spherical symmetry, we put Dirichlet boundary condition on the sphere $\mathcal S_R$ at $r=R$, so that $\Gamma$ is a straight cylinder $\mathcal S_R\times [t_i,t_f]$. As we mentioned, we use our gauge freedom to make sure that $g_{tt}(R)$ is always equal to some prescribed value
\begin{equation}
    g_{tt}(R)=e^{2\nu(R)}f(R)=-\norm{t}^2\equiv-\lambda_R^2.
\end{equation}

The proper energy measured by an observer sitting at $\mathcal S_R$ is defined with respect to its proper time $\dd\tau=\lambda_R\dd t$, so it is related to the canonical energy by a simple redshift factor
\begin{equation}
\begin{split}
\label{eq.energy}
    \mathcal E_R\equiv\frac{E_R}{\lambda_R}=-\sum_k k \hat c_k \int_0^1\dd s\,\pqty{\frac{1-s^2 f(R)}{R^2}}^{k-1} \frac{2\sqrt{f(R)}}{R}\Omega_{d-2} R^{d-2}.
\end{split}
\end{equation}
Analogously, other proper quantities, such as proper angular momentum (which vanishes for this class of metrics) and proper pressure, will be obtained by rescaling.

In order to extract intrinsic quantities, we also need to get rid of finite-size effects by sending the boundary to infinity. However, quantities such as \eqref{eq.energy} generically diverge if we send $R\to\infty$. To get around this problem, we consider the same quantity, but evaluated on a reference background metric with the same form as \eqref{eq.ansatz} and the same boundary conditions, but whose $f\equiv f_{bg}$ function is determined by the field equation \eqref{eq.algfieldeq} with $M=0$. Subtracting the proper energy computed with the background metric from the result \eqref{eq.energy}, we obtain a quantity which is finite in the limit $R\to\infty$. This regularization procedure is called \textit{background subtraction}. Since the background has no mass source, it is natural to expect the difference to be yield to the source mass. Explicit computation gives
\begin{equation}
\label{eq.admmass}
\begin{split}
    \mathcal E_{\infty}^{reg}=&\lim_{R\to \infty}(\mathcal E_R-\mathcal E^{bg}_R)=
    \lim_{R\to \infty}\frac{M}{\sqrt{f_{bg}(R)}},
\end{split}
\end{equation}
where we used the fact that, at large distances, the effect of a massive source decays as $\Delta f\equiv f-f_{bg}\propto 1/R^{d-3}$. This result tells us that, for asymptotically flat spacetimes\,---\,where the reference background is Minkowski ($f_{bg}=1$)\,---\,the energy of the source indeed corresponds to the ADM mass.

If the background solution is not Minkowski, but has constant curvature $\alpha$, then $f_{bg}=1-\alpha r^2$, which corresponds to (anti-)de Sitter for $\alpha>0$ ($\alpha<0$). Correspondingly, the full solution is asymptotically (anti-)de Sitter
\begin{equation}
    f(r)\approx 1-\alpha r^2-\frac{M}{\hat c_1\Omega_{d-2}\,r^{d-3}}+\mathcal O\pqty{\frac{1}{r^{d-2}}}.
\end{equation}
In the asymptotically AdS case, the proper energy measured by an asymptotic observer vanishes because of the infinite redshift $\sqrt{f_{bg}}\sim R/\ell_{AdS}$ (where $\ell_{AdS}$ is the AdS radius). It makes sense to identify the black hole mass as the constant at the numerator, or, in other words, as the asymptotic value of the (canonical) energy $\sqrt{f_{bg}(R)}\mathcal E_R^{reg}$, with the caveat that no observer measures such a mass. In the asymptotically dS case instead, it is impossible to remove finite size effects because the solution has a cosmological horizon (also present in the background solution) outside which there are no stationary observers to refer to.

\subsection{Smarr formula}
\label{s.smarr}

Starting from \eqref{eq.finitesizesmarr} and guided by Equation~\eqref{eq.admmass}, we apply the same procedure to each term: we divide by $\lambda_R$ to get the corresponding proper quantity, subtract the background and finally take the $R\to\infty$ limit after multiplying by $\sqrt{f_{bg}(R)}$. In this way, we arrive at the Smarr formula
\begin{equation}
\label{eq.smarr}
   (d-3)M=(d-2)\Omega^i_H {\mathcal J}^i+(d-2)T_H S+2\sum_k(k-1)c_k\psi_k-(d-2)\sigma A,
\end{equation}
where we introduced the regularized quantities\footnote{We used the same symbol to indicate both the proper surface tension and the tension term in the generalized Brown-York tensor in \eqref{eq.surfacetensionterm}. Indeed, for a spherically symmetric solution where $\sigma$ only depends on $R$, one can easily see that $\Sigma=\lambda_R\sigma$.\label{ft.surfacetension}}
\begin{gather}
    M=\lim_{R\to\infty}\sqrt{f_{bg}(R)}\pqty{\frac{E}{\lambda_R}-\frac{E}{\lambda_R}\eval_{M=0}},\\
    \psi_k=\lim_{R\to\infty}\sqrt{f_{bg}(R)}\pqty{\frac{\Psi_k}{\lambda_R}-\frac{\Psi_k}{\lambda_R}\eval_{M=0}},\\
    \sigma=\lim_{R\to\infty}\sqrt{f_{bg}(R)}\pqty{\frac{\tilde\Sigma}{\lambda_R}-\frac{\tilde\Sigma}{\lambda_R}\eval_{M=0}},
\end{gather}
and the Hawking temperature
\begin{equation}
    T_H=\lim_{R\to\infty}\frac{\sqrt{f_{bg}(R)}}{\lambda_R}\frac{\kappa}{2\pi}=\lim_{R\to\infty}\,\sqrt{f_{bg}(R)}T.
\end{equation}

\subsection{Comparison}

We now comment on the form of the thermodynamic potentials \eqref{eq.thermopotentials}, comparing $\psi_k$ with the expression (37) of Ref.~\cite{Kastor:2010gq}, that we report here:
\begin{equation}
    \Psi^{(k)}_{KRT}=B^{(k)}+2\kappa A^{(k)}+\Theta^{(k)}. 
\end{equation}
In this expression, the authors of \cite{Kastor:2010gq} reconstruct the thermodynamical potential conjugated to the ratio $c_k/c_1$ (there called $b_k$) in terms of three quantities:
\begin{enumerate}
    \item $B^{(k)}$, defined as the boundary integral of the codimension 2 form that measures the difference of (the $k$-th term contribution to) the on-shell Hamiltonian density between two nearby solutions, one with a black hole and one without. Notice that the authors assume to be in asymptotically AdS spacetime, hence the solution wiht no black hole is just pure AdS;
    
    \item $A^{(k)}$, which is just the integral of $k\mathbf{L}_0^{k-1}$ over an horizon cross-section;
    
    \item $\Theta^{(k)}$, defined as the difference between the Killing-Lovelock potentials integrated on the sphere at infinity $\mathcal S_\infty$ and on an horizon cross-section. Since the integral at $\mathcal S_\infty$ makes the difference diverge, they regulate $\Theta^{(k)}$ by background subtraction, that is subtracting the Killing-Lovelock potential integrated on the $\mathcal S_\infty$ of the AdS background.
\end{enumerate}

For the sake of comparison, we assume that we are dealing with a black hole in AdS spacetime, we take $\Sigma$ to go from $\mathcal S$ to $\mathcal B$ in evaluating the Noether charge of the solution\footnote{In our treatment, we chose the bifurcation surface $\mathcal B$ as a preferred horizon cross-section, but, due to time translation invariance, any other cross-section would give the same result.}, but we extend $\Sigma$ down to the origin in the reference AdS background since there is no horizon.
\begin{gather}
\label{eq.solthermopotential}
    \Psi_k=\int_{\mathcal S} (\mathbf{q}^{(k)}_2[\xi]-\mathbf{J}^{(k)}_2[\xi])-\int_{\mathcal B} \mathbf{q}^{(k)}_2[\xi]-\int_\Sigma \mathbf{K}^{(k)}_1[\xi],\\
\label{eq.bgthermopotential}
    \Psi^{bg}_k=\int_{\mathcal S^{AdS}} (\mathbf{q}^{(k)}_2[\xi]-\mathbf{J}^{(k)}_2[\xi])+0-\int_\Sigma^{AdS} \mathbf{K}^{(k)}_1[\xi].
\end{gather}
Let us start from the end to show that these three pieces, when subtracted one by one, exactly correspond to the three pieces in $\Psi^{(k)}_{KRT}$.
\begin{etaremune}
   \item As we discussed in the previous section, the bulk integral of the constraint currents $\mathbf{K}_1^{(k)}[\xi]$ can be replaced by the integrals of the Killing-Lovelock potentials at $\mathcal S_R$ and $\mathcal B$ when there is a black hole, and at $\mathcal S_R$ only when there is not. We thus recognize that the difference between the last two integrals in \eqref{eq.solthermopotential} and \eqref{eq.bgthermopotential} generalizes $\Theta^{(k)}$ for finite-distance boundaries, while recovering it when $R\to\infty$.
    \item Second, the term at the horizon in \eqref{eq.solthermopotential} clearly coincides with $2\kappa A^{(k)}$, as one can check using \eqref{eq.noethercharge}.
    \item Finally, let us observe that, in \cite{Kastor:2010gq}, the Hamiltonian density is identified with the constraint current. Hence, a variation of the latter generates a flow only with respect to the bulk part of the symplectic current \eqref{eq.symcurrent}:
    \begin{equation}
    \delta \mathbf{K}_1[\xi]= -\delta\bm\Theta_1(g,\delta_\xi g)+\delta_\xi\bm\Theta_1(g,\delta g)+\dd(\delta\mathbf{q}_2[\xi]-\xi\cdot\bm\Theta_1(g,\delta g)).
    \end{equation}
    Therefore, the aforementioned codimension $2$ form which measures the difference $\delta \mathbf{K}_1[\xi]$ between two nearby solutions is $\delta\mathbf{q}_2[\xi]-\xi\cdot\bm\Theta_1(g,\delta g)$. Whereas the first term is exact, the second is not, and it easy to see that one has to require the existence of the famous $\mathbf{B}_2(\xi)$ form (remember that it was defined by $\delta \mathbf B_2(\xi)=\xi\cdot \bm\Theta_1$) in order to get an integrable result.
    As we discussed at the end of Section \ref{ss.firstlaw}, these terms (one for each $k$) are nothing but the boundary current $\mathbf{J}_2^{(k)}[\xi]$. With this identification, the $B^{(k)}$ term, which is not exact as we would expect, is reconstructed\,---\,and generalized to finite-distance boundaries\,---\,by the difference between the first terms in \eqref{eq.solthermopotential} and \eqref{eq.bgthermopotential}.
\end{etaremune}

\section{Canonical ensemble derivation}
\label{s.ced}

In this section, we give an alternative derivation of the formulas we have obtained in this paper from the usual Euclidean path integral approach \cite{Gibbons:1994cg}. Replacing $t\to i\tau$ in \eqref{eq.ansatz}, we readily obtain the ansatz for static and spherically symmetric saddle-points
\begin{equation}
\label{eq.euclideanansatz}
    \dd s^2=e^{2\nu(r)} f(r)\dd \tau^2+\frac{\dd r^2}{f(r)}+r^2\dd\Omega_{d-2}^2.
\end{equation}
In the classical limit, we approximate the result of the full path integral with the value of its integrand $e^{-I_E}$ on a saddle-point. In this approximation, we can identify the on-shell Euclidean action as $\beta \mathcal F$, where $\mathcal F$ is the Helmholtz free energy of a canonical thermodynamical system.
The reason why this analogy works is that Dirichlet boundary conditions on $\Gamma=S^1\times S^{d-2}$ amount to fixing the size of the system (through the boundary sphere line element $R^2\dd\Omega^2_{d-2}$) and the temperature, given by the inverse proper length of the Euclidean circle $\beta=\lambda_R \Delta\tau$.

Among the possible values of $\beta$, only one is compatible with the presence of an horizon and regularity, because the near horizon line element is that of a cone. To show this, we use $\rho$ to measure the proper length from the horizon, and get the line element
\begin{equation}
    \dd s^2\approx \frac{e^{2\nu(r_H)}}{4}f'(r_H)^2\rho^2\dd\tau^2+\dd\rho^2+r_H^2\dd\Omega_{d-2}^2
\end{equation}
which is indeed a cone times a $(d-2)$-sphere. To avoid the singularity at the tip of the cone (i.e. to flatten the cone), we must impose that the angular coordinate has period $2\pi$. This fixes the value of $\beta$\footnote{We used the relation 
$\lambda_R=\sqrt{f(R)}e^{\nu(R)}$ and the on-shell constancy of $\nu$.}
\begin{equation}
\label{eq.noconicalsing}
    \frac{e^{\nu(r_H)}}{2}f'(r_H)\Delta\tau=2\pi\implies \beta=\frac{4\pi\sqrt{f(R)}}{f'(r_H)}.
\end{equation}
The resulting topology is that of a cigar $D^2\times S^{d-2}$, whose boundary is $S^1\times S^{d-2}$.

The total on-shell action, expressed as a function of $\beta$ and $R$ is
\begin{equation}
\begin{split}
    I_E[\beta,R]=&\,2\pi\sum_k \frac{2k \hat c_k}{d-2k}\pqty{\frac{1}{r_H^2}}^{k-1}\Omega_{d-2}r_H^{d-2}+\\
    &+\beta\sum_k k\hat c_k\int_0^1 \dd s\pqty{\frac{1-s^2 f_R}{R^2}}^{k-1}\frac{2 \sqrt{f_R}}{R}\Omega_{d-2} R^{d-2},
\end{split}
\end{equation}
where $f_R\equiv f(R)$ and $r_H$ are intended as functions of $\beta$ and $R$. The action comprises two pieces, one evaluated at the horizon and one at the boundary sphere. Notice that, in the former, the time interval combines precisely with the $f'(r_H)$ factor to give $2\pi$. Following~\cite{Bradem:1990}, we lay out the procedure to obtain the expressions for $f_R$ and $r_H$.
\begin{enumerate}
    \item Solve the field equations relating the value of the integrated constraint at different radii
    \begin{equation}
        \label{eq.fieldequation}
    \sum_k \hat c_k \pqty{\frac{1}{r_H^2}}^k r_H^{d-1}=\sum_k \hat c_k \pqty{\frac{1-f_R}{R^2}}^k R^{d-1},
    \end{equation}
    to get $f_R(r_H,R)$. Taking a derivative with respect to $r_H$, we get
    \begin{equation}
    \pdv{f_R}{\,r_H}=-\frac{\sum_k (d-2k-1)\hat c_k\pqty{\frac{1}{r_H^2}}^k}{\sum_{k'} {k'} \hat c_{k'} \pqty{\frac{1-f_R}{R^2}}^{k'-1}}\frac{r_H^{d-2}}{R^{d-3}}.
    \end{equation}
    \item Remove the total dependence of the action on $r_H$ using the last formula to compute the dependence implicit in $f_R$. Setting $\dd I/\dd r_H=0$ gives
    \begin{equation}
    \label{eq.extremizinghorizon}
        \sum_k 2k \hat c_k\pqty{\frac{1}{r_H^2}}^{k-1}=\frac{\beta}{2\pi}\frac{r_H}{\sqrt{f_R}}\sum_k (d-2k-1)\hat c_k\pqty{\frac{1}{r_H^2}}^k
    \end{equation}
    which can be solved to express $r_H$ in terms of $\beta$ and $R$ once we plug the solution of \eqref{eq.fieldequation} in $f_R$. Notice that this equation is compatible with the value of $\beta$ that removes the conical singularity.
    \item Insert the function $r_H(\beta,R)$ in $f_R(r_H,R)$ to get the function $f_R(\beta,R)$. This procedure does not lead to a unique solution, as expected from the existence of multiple vacua in a generic LL theory.
\end{enumerate}

Once we have the free energy, we can extract thermodynamic quantities through the usual relations\footnote{The dependence on $\beta$ from $r_H$ cancels because of the extremization condition \eqref{eq.extremizinghorizon}.}
\begin{align}
    \mathcal E(\beta,R)&=\pdv{I_E}{\beta}=-\sum_k k\hat c_k\int_0^1 \dd s\pqty{\frac{1-s^2 f_R}{R^2}}^{k-1}\frac{2 \sqrt{f_R}}{R}\Omega_{d-2} R^{d-2},\\
    S(\beta,R)&=\beta\pdv{I_E}{\beta}-I_E=2\pi\sum_k \frac{2k \hat c_k}{d-2k}\pqty{\frac{1}{r_H^2}}^{k-1}\Omega_{d-2}r_H^{d-2},\\
    \sigma(\beta,R)&=-\frac{1}{\beta}\pdv{I_E}{A}
\end{align}
with $A=\Omega_{d-2} R^{d-2}$ being the surface area of the spherical box. The first two quantities coincide respectively with the energy $\mathcal E_{R}$ computed in \eqref{eq.energy} and the entropy that we identified in \eqref{eq.waldentropy}. The last quantity, at least for the specific examples that we will present, coincide with $\Sigma/\lambda_R$. With these definitions, we can justify the thermodynamical description
\begin{equation}
    I_E=\beta \mathcal E-S=\beta \mathcal F.
\end{equation}

\subsection{Black hole thermodynamics in 4d}
\label{ss.bh4d}

We specialize the previous analysis for the well-known case of GR with a cosmological constant. We show that our prescription for the thermodynamical potential gives the correct Smarr formula as in~\cite{Kastor:2008xb}.
If we express the ratio of the coefficients in terms of the cosmological constant $c_0/c_1=-2\Lambda$, we write the solution of \eqref{eq.fieldequation} as
\begin{equation}
    f_R=1-\frac{r_H}{R}+\frac{\Lambda}{3}\frac{R^3-r_H^3}{R}.
\end{equation}
We will focus only on the case $\Lambda\le 0$ because it allows us to take the $R\to\infty$ limit in a meaningful way. One can still extend our finite-radius results to the $\Lambda>0$ (de Sitter) case. The position of the black hole horizon is related to the inverse temperature by solving \eqref{eq.extremizinghorizon}
\begin{equation}
    \beta=4\pi r_H\pqty{1-\Lambda r_H^2}^{-1}\pqty{1-\frac{r_H}{R}-\frac{\Lambda}{3}\frac{R^3-r_H^3}{R}}^{1/2}.
\end{equation}
It is interesting to compare this with the usual expression in terms of the ADM mass
\begin{equation}
\label{eq.adsschwarzschildradius}
    3r_H-\Lambda r_H^3=6\,G_N M.
\end{equation}
These last two equations tell us that, if we send $R\to\infty$ at fixed $\beta$ in an asymptotically AdS spacetime, the ADM mass goes to $0$.
We can compute the on-shell value of the Euclidean action to get the free energy
\begin{equation}
    I_E[\beta, R]=-\frac{\pi r_H^2}{G_N}-\frac{\beta R}{G_N}\sqrt{1-\frac{r_H}{R}-\frac{\Lambda}{3}\frac{R^3-r_H^3}{R}}.
\end{equation}
Perhaps unsurprisingly for the reader, the energy, entropy and surface tension that we obtain are
\begin{gather}
    \mathcal E=-\frac{R}{G_N}\sqrt{1-\frac{2 G_N M}{R}-\frac{\Lambda}{3}R^2},\\
    S=\frac{A_H}{4G},\\
    \sigma=-\frac{1}{8\pi R}\frac{1}{G_N\mathcal E/R}\pqty{\frac{M}{R}+\frac{G_N\mathcal E^2}{R^2}-\frac{\Lambda R^2}{3G_N}}.
\end{gather}
These quantities satisfy the physical version of the first law $\delta\mathcal E=T\delta S-\sigma \delta A$.

The only non-trivial thermodynamical potential in this case is the one conjugate to the cosmological constant, which is basically the volume of the hypersurface $\Sigma$, up to a constant factor
\begin{equation}
    \Psi_0=\int_\Sigma \mathbf{J}_1^0=\frac{\lambda_R}{\sqrt{f_R}}\frac{4\pi}{3} (R^3-r_H^3).
\end{equation}
Even for finite size boundaries, we can check that \eqref{eq.finitesizesmarr} is satisfied by
\begin{equation}
    E=2\,\lambda_R T S-2\tilde\Sigma A+2P_\Lambda \Psi_0,
\end{equation}
where we introduced the cosmological pressure $-\Lambda/(8\pi G_N)$. Notice that the temperature $T=1/\beta$ is multiplied by the norm of the Killing vector $\norm{\xi}=\lambda_R$.
From here, we finally recover the Smarr formula using background subtraction, as described in the previous section
\begin{equation}
    M=2\,T_H S-2P_\Lambda V_{eff}.
\end{equation}
While the formula looks familiar, let us pause to describe what happened. In order to recover the mass, we divided both sides of the equation by $\lambda_R$, which was present explicitly in the temperature-entropy term and implicitly in $\tilde\Sigma=\lambda_R\sigma$ (cfr. footnote \ref{ft.surfacetension}) and $\Psi_0$. When we subtracted against the background, the $TS$ term did not get regularized, but we needed to multiply by $\sqrt{f_{bg}(R)}$ to get a sensible result in the $R\to\infty$ limit. Doing so, we recover the correct Hawking temperature
\begin{equation}
    T_H=\lim_{R\to\infty} \sqrt{1-\frac{\Lambda}{3} R^2}\, T=\frac{1-\Lambda r_H
    ^2}{4\pi r_H}.
\end{equation}
On the other hand, both the surface tension and the cosmological pressure term contained background pieces (which dropped from the subtraction) and neither of them vanished in the $R\to\infty$ limit. We wrote the result in terms of the effective black hole volume $V_{\rm eff}\equiv(4\pi/3)r_H^3$, which coincide with the volume of a flat ball with the same radius as the horizon. Let us notice that, in this framework where we keep track of boundary terms, we obtained $-2P_\Lambda V_{\rm eff}$ as a combination of effects between the cosmological and the thermodynamical pressure.

\subsection{Black hole thermodynamics in 5d}
\label{ss.bh5d}

In five dimensions, we consider Einstein-Gauss-Bonnet theory, which is obtained for $c_0=0$ and $c_1\ne 0$, $c_2\ne 0$. In this theory, there are two coexisting vacua, one of which is flat. We choose to present here solutions that approach the flat vacuum asymptotically, since the other vacuum is commonly believed to be pathological \cite{Boulware:1985wk}. If we write $c_2/c_1$ as $\alpha_{GB}/4$, the solution of \eqref{eq.fieldequation} is
\begin{gather}
    f_R=1+\frac{R^2}{\alpha_{GB}}-\frac{1}{\alpha_{GB}}\sqrt{R^4+2r_H^2\alpha_{GB}+\alpha_{GB}^2}\approx 1-\frac{2r_H^2+\alpha_{GB}}{2R^2},
\end{gather}
where, in terms of the ADM mass
\begin{equation}
    r_H^2=\frac{8G_N M}{3\pi}-\frac{\alpha_{GB}}{2}.
\end{equation}
Since we want $r_H$ to be a zero of the $f$ function, we require $r_H^2+\alpha_{GB}\ge0$ in addition to $2r_H^2+\alpha_{GB}\propto G_N M\ge0$. The relation between the inverse temperature and the horizon radius is
\begin{equation}
    \beta=\frac{2\pi}{r_H}(r_H^2+\alpha_{GB})\sqrt{1+\frac{R^2}{\alpha_{GB}}-\frac{1}{\alpha_{GB}}\sqrt{R^4+2r_H^2\alpha_{GB}+\alpha_{GB}^2}}.
\end{equation}
which can be analytically solved in the $R\to\infty$ limit
\begin{equation}
    r_H=\frac{1}{4\pi}\pqty{\beta+\sqrt{\beta^2-16\pi^2\alpha_{GB}}}+\mathcal O\pqty{\frac{1}{R^2}}.
\end{equation}

Except for the entropy, the thermodynamical quantities do not have a particularly nice form, so we only write down their large $R$ expansion
\begin{gather}
    \mathcal E\approx  M-\frac{\pi\alpha_{GB}}{2G_N}-\frac{3\pi R^2}{4 G_N}+\mathcal O\pqty{\frac{1}{R^2}},\\
    S=\frac{A_H}{4 G_N}+\frac{3\pi^2\alpha_{GB}r_H}{2G_N},\\
    \sigma\approx\frac{1}{4\pi G_N R}+\mathcal O\pqty{\frac{1}{R^5}}.
\end{gather}
In this case, the thermodynamical potential of interest is $\Psi_2$, which can be computed exactly, but we only report in the asymptotic expansion
\begin{equation}
    \Psi_2\approx 4\pi^2\lambda_R \frac{17r_H^2+5\alpha_{GB}}{r_H^2+\alpha_{GB}}+\mathcal O\pqty{\frac{1}{R^2}}.
\end{equation}
To obtain the Smarr formula, we have to confront the relation
\begin{equation}
    2\mathcal E=3T S-\frac{\alpha_{GB}}{32\pi G_N}\frac{\Psi_2}{\lambda_R}-3\sigma A
\end{equation}
with that of a fixed background and then take the $R\to\infty$ limit. In Section \ref{ss.recoveringsi}, we chose the reference background to be given by the solution of \eqref{eq.algfieldeq} with $M=0$. However, in this case, $r_H$ does not vanish when $M\to 0$ and, consequently, neither does the entropy. To get a sensible result, we choose to regularize against the background generated by the constant mass $M_{bg}\equiv3\pi\alpha_{GB}/16 G_N$, for which the horizon radius does vanish
\begin{equation}
    2M-2M_{bg}=3T_H S-\frac{3\pi\alpha_{GB}}{2 G_N}\frac{r_H^2}{r_H^2+\alpha_{GB}}.
\end{equation}
In this case, because of asymptotic flatness, there was no need to multiply by $\sqrt{f_{bg}(R)}$ to get a non-zero result, and the Hawking temperature that we get coincide with the large $R$ limit of $1/\beta$
\begin{equation}
    T_H=\frac{r_H}{2\pi(r_H^2+\alpha_{GB})}.
\end{equation}

\subsection{Area law}

It has been suggested that one should be able to recover the area law for the entropy in modified theories of gravity such as LL~\cite{Kubiznak:2016qmn}. To do so, one needs to cancel the contribution to the entropy coming from higher-order terms
\begin{equation}
    S=\frac{A_H}{4G}+2\pi\sum_{k>1}\frac{2k \hat c_k}{d-2k}\pqty{\frac{1}{r_H^2}}^{k-1}\Omega_{d-2}r_H^{d-2}
\end{equation}
Before dealing with modifications from the area law, we notice that, in the critical dimension, the ratio $\hat c_k$ to $d-2k$ is finite. This means that, even if the Lovelock term does not contribute to the equations of motion, it appears to affect the entropy. This paradoxical conclusion is solved by noticing that the contribution is a constant. In fact, the term comes from integrating the Lovelock Lagrangian of order $d/2-1$, which is topological for a $(d-2)$-dimensional manifold (the horizon). The result is proportional to the Euler characteristic of that manifold, which is $2$ for an even dimensional sphere.

Furthermore, since $\mathcal G^{(k)}$ is identically $0$ in $d=2k$, one can see that the corresponding Lovelock Lagrangian is a total derivative also off-shell \cite{Padmanabhan:2013xyr}. Let us call $\mathbf{T}^{(k)}_1$ the codimension $1$ form such that $\mathbf{L}^{(k)}_0\overset{d=2k}{=}\dd \mathbf{T}_1$. In this case, one can clearly see why the $k=d/2$ term does not contribute to the field equations in any even dimension and can be eliminated from the Lagrangian.
Despite this is not true in other dimensions, nothing prevents us from considering a new pair of Lagrangians $(\mathbf{L}'_0,\mathbf{L}'_1)$ with $\mathbf{L}'_0=\mathbf{L}'_0-\sum_{k>1} c_k \dd\mathbf{T}^{(k)}_1$ and $\mathbf{L}'_1=\mathbf{L}'_1-\sum_{k>1} c_k j^*_\Gamma \mathbf{T}^{(k)}_1$.
Notably, the new bulk term continues to be irrelevant for the field equations and it does not change the symplectic structure in the Lorentzian setting, as $\bm\Theta_1$ changes by an exact piece and $\bm\Theta_2$ does not change. On the other hand, in the Euclidean setting, we are naturally lead to excise a part of spacetime due to the pinching of the Euclidean time circles at the horizon. 
The on-shell Euclidean action changes as
\begin{equation}
    I_E\to I'_E=I_E+\sum_{k>1} c_k \int_{\mathcal B\times \mathcal S^1} \mathbf{T}_1^{(k)},
    \label{eq.KubMod}
\end{equation}
where the last term is finite because $\mathbf{T}_1^{(k)}$ diverges as $1/\norm{\xi}$ at the horizon, but the Euclidean circle length shrinks to zero at the same rate.
Explicitly, for a spherically symmetric and static spacetime, one gets
\begin{equation}
    I_E\to I'_E=I_E+2\pi\sum_{k>1} \frac{2k \hat c_k}{d-2k} \pqty{\frac{1}{r_H^2}}^{k-1}\Omega_{d-2}r_H^{d-2},
\end{equation}
so that the corresponding entropy reduces to the Bekenstein law
\begin{equation}
    S\to S'=\frac{A_H}{4G}.
\end{equation}

It thus appear that we can freely change the entropy of the system without altering its physics, but this conclusion seems absurd! We now provide some arguments as to why the previous redefinition of the action is not harmless as it seems.
First of all, we can easily see that there exists a $\mathbf{T}^{(1)}$ too, which, if removed from the bulk Lagrangian, would lead to a system with zero entropy.
Second, modifying the action as suggested requires to partially break covariance, introducing anomalies for diffeomorphisms generated by vector fields that do not commute with $\xi$. While possible, we believe that this should be allowed only at the level of effective descriptions as to not break fundamental covariance.
Furthermore, the black hole boundary conditions that remove the conical singularity seem to be incompatible with the extremization procedure or, alternatively, imposing~\eqref{eq.noconicalsing} makes so that the new action is not stationary against changes in the horizon radius (which is not fixed in the canonical ensemble). Indeed, one can compute that
\begin{equation}
    \fdv{I'_E}{r_H}=-\frac{4\pi\lambda_R}{\beta}\sum_{k\ne 1} k\hat c_k\pqty{\frac{1}{r_H^2}}^{k-1}r_H^{d-3},
\end{equation}
which does not generically vanishes.
Let us make a final comment. In order for the first law to be preserved, entropy cannot be modified unless the Hawking temperature is modified too. It is known that, in Horndeski theories of gravity, the speed of gravitons is affected by the non-minimal coupling between the scalar field and the background metric, entailing the need to modify the temperature which enters in the first law. 

The authors of~\cite{Kubiznak:2016qmn} suggests a similar mechanism as a reason to justify the needed modification of the temperature ensuing the entropy change. Due to the intrinsic non-linearity of LL theories (remember that only in GR Lagrangian, metric second derivatives appear linearly \cite{Deruelle:2003ck}), one could expect that gravitons on a background propagate with respect to an effective metric which is not the background, and thus move at a speed which is different from the speed of light. While this is certainly true, it has been shown though that radially propagating gravitons move exactly at $c$, thus giving no reason to modify the temperature in this context \cite{Reall:2014pwa, Brustein18}.

While the idea of having an entropy which is manifestly positive definite is appealing, many results have been achieved regarding the validity of the second law for the entropy in LL theories \cite{Kolekar:2012tq, Sarkar:2013swa} and generally covariant theories in general \cite{Wall:2015raa, Wall:2024lbd} and this reduces the problem to have the entropy start out positive. 

\section{Discussion}

In this study, we presented a novel prescription for determining the thermodynamic potentials associated to the couplings of a Lanczos-Lovelock theory. These potentials emerge when there are multiple couplings in the theory, as in this case their ratios provide intrinsic length scales which, in turn, break scale invariance. Since scale invariance is crucial for deriving the Smarr formula, one has to resort to an extended framework of thermodynamics where couplings are treated as fields, i.e.~they can vary, and hence require the existence of associated thermodynamic potentials.

Our prescription came out naturally from a quick derivation of the first law of (extended) thermodynamics within the covariant phase space formalism. Notably, it differs from previous ones, given e.g.~in \cite{Kastor:2010gq}, by enabling the construction of potentials which are finite without reference to any background.
In fact, throughout the paper, we consistently worked with finite-size systems contained within Dirichlet walls: it is only when one desires to remove the boundary that quantities might diverge. Handling such limit required the inclusion of a surface tension term, obviously absent when dealing with unconfined systems, which end up shedding some more light on the way the $R\to\infty$ limit is realized. For example, we showed in Section~\ref{ss.bh4d} that this term does not vanish when $R\to\infty$, but instead combines with the term associated with the cosmological constant variation to yield the well-known ``cosmological pressure" term of the extended black hole thermodynamics \cite{Kastor:2008xb}.

Another desirable feature of the here presented approach is related to the fact that, as demonstrated in \cite{Harlow:2015lma}, the covariant phase space algorithm provides an unambiguous definition of the energy without requiring any ``integrability condition'' for the symplectic potential at the boundary. This same virtue imbues our prescription for the thermodynamic potentials.

For what concerns the future developments of this investigation, we start by remarking that our derivation of the Smarr formula, Eq.~\eqref{eq.finitesizesmarr}, albeit being very general, still rests on the assumption that both the black hole and the perturbations are stationary. Since much work has been done in the non-stationary case \cite{Bhattacharjee:2015yaa, Mishra:2017sqs, Davies:2022xdq, Visser:2024pwz, Hollands:2024vbe, Wall:2024lbd}, it would be  important to understand whether our formalism can be extended to this case and, if so, extract and compare the results with the aforementioned literature.

In addition to the above point, let us stress that the main limitation of our work lies perhaps in the derivation of the Smarr formula, where we focused solely on a narrow (even if interesting) class of black holes: static, spherically symmetric, and chargeless, i.e. Schwarzschild-like. This was required in order to have an explicit relation between the energy measured by a boundary observer and the ADM mass, which is the most relevant quantity entering the Smarr formula in $d>3$.
We anticipate that extending our approach to a more general class of black holes should not be hindered by the shortage of exact analytic solutions because the relation we need relies on the $R\to\infty$ behaviour of the solution, where its bulk properties become irrelevant.

Generalizing our findings to include non-spherical horizon topologies and charged black holes \cite{Hull:2021bry, Wheeler:1985nh} should not present more difficulties than the ones we discussed here. See e.g. \cite{Cai:2003kt} for the standard derivation of the Smarr formula that extend to these cases.
On the other hand, the generalization to rotating black holes requires more care, particularly in disentangling the contribution of the ADM mass to the canonical energy. Moreover, even if we were successful, validating our results would be challenging due to the difficulty in finding exact analytic solutions for rotating black holes \cite{Kim2008, Brihaye:2008kh, Suzuki&Tomizawa2022}. 
We leave this investigation for the future, hoping that the present work will stimulate further studies and discussions in this and other directions.

\section*{Acknowledgements}

G. Neri expresses its gratitude to Simon Langenscheidt for useful and clarifying conversations about the covariant phase space formalism, and for proofreading the draft of this paper.

\appendix
\section{Boundary Lagrangian and corner terms}
\label{app.blct}

In this appendix, we show that the boundary Lagrangian \eqref{eq.boundlag} is the correct one to impose Dirichlet boundary conditions for the metric on $\Gamma$, that is $\delta q\equiv \delta(j^*_\Gamma g)=0$. In other words, we will show the validity of \eqref{eq.boundlagvar}. In doing so, we also derive the expressions of the generalized Brown-York tensor and the corner symplectic potential. Throughout this appendix, we do not specify the LL index $k$ on every object since the result holds separately for each $k$.

To shorten the derivation, we introduce a tensor on $\Gamma$ inspired by the Lovelock tensor 
\begin{equation}
    p^{iab}_{jcd}(s)=\frac{k}{2^{k-1}}\delta^{iaba_3b_3\dots a_{k}b_{k}}_{jcdc_3d_3\dots c_{k}d_{k}}\prod_{i=3}^{k} \pqty{R^{c_i d_i}_{a_i b_i}-2s^2  K^{c_i}_{a_i}  K^{d_i}_{b_i}}.
\end{equation}
Using this tensor, the boundary Lagrangian (at least for $k>1$) reads\footnote{For $k=0$ and $k=1$, the boundary Lagrangian is respectively $0$ and $2K$.}
\begin{equation}
    L_1=2\int_0^1\dd s\, p^{iab}_{jcd}(s)K^j_i\pqty{R^{c d}_{a b}-2s^2  K^{c}_{a}  K^{d}_{b}}.
\end{equation}
Let us notice that, upon variation, we can express the result in terms of a $\delta R$ and a $\delta K$ piece. The remarkable property is that the latter multiplies a term that can be integrated exactly: in fact, there are in total $2k-2$ (from the variation of the interpolating curvature) $+1$ (from the variation of the extrinsic curvature) $\delta K$ terms which multiply a product of two $K$'s and just one $\delta K$ which multiply an intrinsic curvature term. Thanks to the antisymmetry of the generalized Kronecker delta, all these terms effectively commute and we can use the inverse product rule to get
\begin{equation}
\begin{split}
    &\delta K^j_i(R^{cd}_{ab}-2s^2(2k-1)  K^{c}_{a}  K^{d}_{b})\delta^{iaba_3b_3\dots a_{k}b_{k}}_{jcdc_3d_3\dots c_{k}d_{k}}\prod_{i=3}^{k} \pqty{R^{c_i d_i}_{a_i b_i}-2s^2  K^{c_i}_{a_i}  K^{d_i}_{b_i}}=\\
    &\delta K^j_i\dv{}{s}\pqty{s\delta^{ia_2b_2a_3b_3\dots a_{k}b_{k}}_{jc_2d_2c_3d_3\dots c_{k}d_{k}}\prod_{i=2}^{k} \pqty{R^{c_i d_i}_{a_i b_i}-2s^2  K^{c_i}_{a_i}  K^{d_i}_{b_i}}}.
    \end{split}
\end{equation}
The variation of  $ L_1$ is thus
\begin{equation}
\label{eq.boundaryterm1}
    \delta L_1=2p^{iab}_{jcd}(1)\delta K^j_i\pqty{R^{c d}_{a b}-2 K^{c}_{a}  K^{d}_{b}}+ 2(k-1)\int_0^1\dd s\, p^{iab}_{jcd}(s) K^j_i\delta R^{cd}_{ab}.
\end{equation}
Before going on, let us write down the explicit form of the extrinsic curvature variation
\begin{equation}
\label{eq.extcurvaturevar}
    \delta K_{\mu\nu}=n\cdot \delta n\, K_{\mu\nu}+\not\!\delta c_\mu n_\nu+\not\!\delta c_\nu n_\mu-(q^\lambda_\mu q^\beta_\nu  n^\alpha+q^\alpha_\mu q^\lambda_\nu  n^\beta-q^\beta_\mu q^\alpha_\nu  n^\lambda)\nabla_\lambda\delta g_{\alpha\beta}
\end{equation}
where, for notation simplicity, we introduced the tangential tensor $\not\!\delta c_\mu=\delta g_{\alpha\beta}n^\alpha q^\beta_\mu$ and we used the same symbol as the induced metric $q$ to denote the projector on $\Gamma$.

The other term that we need is the pull-back to $\Gamma$ of the boundary symplectic potential $j^*_\Gamma\bm\Theta_1=(n\cdot \Theta_1)\bm\epsilon_\Gamma$.
Since the Lovelock tensor is antisymmetric on both upper and lower indices, we can project at most one index from each pair along $n$. The contraction $n\cdot\Theta_1$ already forces our hand on the upper pair: the second index must be projected along $\Gamma$. Accordingly, let us introduce the tangential tensor $\pi^\alpha_\beta \equiv P^{\alpha\nu}_{\beta\sigma}n_\nu n^\sigma$ and the mixed tensor $\sigma^\alpha{}_{\beta\gamma}\equiv P^{\alpha\nu}_{\beta\gamma}n_\nu$, antisymmetric on its last two indices, for which
\begin{equation}
    \sigma^\alpha{}_{\beta\gamma}=2\pi^\alpha_{[\beta}n_{\gamma]}+\bar\sigma^\alpha{}_{\beta\gamma},\quad\text{with}\quad n_\alpha\bar\sigma^\alpha{}_{\beta\gamma}=n^\beta\bar\sigma^\alpha{}_{\beta\gamma}=n^\gamma \bar\sigma^\alpha{}_{\beta\gamma}=0,
\end{equation}
In terms of these tensors, the projected boundary symplectic potential reads
\begin{equation}
\begin{split}
    & n_\mu \Theta_1^\mu=2\bar  \sigma^{\mu\nu\rho}\nabla_\nu \delta g_{\mu\rho}+2\pi^{\mu\nu}n^\rho\nabla_\nu \delta g_{\mu\rho}-2\pi^{\mu\rho}n^\nu\nabla_\nu \delta g_{\mu\rho}.
\end{split}
\end{equation}
Let us call these three pieces $(a)$, $(b)$ and $(c)$.
Integrating by parts, we are able to write
\begin{equation}
\begin{split}
    (a): &-2\delta g_{\mu\rho}\nabla_\nu \bar\sigma^{\mu\nu\rho}+2\nabla_\nu(\bar \sigma^{\mu\nu\rho}\delta g_{\mu\rho})=\\
    &=-2\delta q_{bc}D_a\bar\sigma^{bac}+2\!\not\!\delta c_a \bar\sigma^{bca}K_{bc}+2D_a(\bar\sigma^{bac}\delta q_{bc}).
\end{split}
\end{equation}
Then, we use \eqref{eq.extcurvaturevar} to rewrite the third term
\begin{equation}
    (c):-4\pi^{ca}\delta K_{ac}+4\pi^{i}_{j}K^j_i n\cdot \delta n-2\pi^{\nu\sigma} n^\mu \nabla_\nu \delta g_{\mu\rho}-2\pi^{\mu\nu} n^\rho \nabla_\nu \delta g_{\mu\rho}.
\end{equation}
Notice that the last term of this expression we just obtained cancels $(b)$ exactly. The next to last, on the other hand, can be expanded as
\begin{equation}
\begin{split}
    -2\pi^{\nu\sigma} n^\mu \nabla_\nu \delta g_{\mu\rho}&=-2q^\mu_\nu\nabla_\mu(\pi^{\nu}_{\rho}\!\not\!\delta c^\rho)+2q^\mu_\nu\!\not\!\delta c^\rho\nabla_\mu \pi^{\nu}_{\rho}+2\pi^{cb}K^a_c \delta q_{ab}-4\pi^i_j K^j_i n\cdot \delta n=\\&=-2D_a(\pi^{ab}\!\not\!\delta c_b)+2\!\not\!\delta c_a D_b \pi^{ba}+2\pi^{cb}K^a_c \delta q_{ab}-4\pi^i_j K^j_i n\cdot \delta n,
    \end{split}
\end{equation}
When putting these pieces back together, we notice that there are some nice cancellations. In particular, all the terms involving $n\cdot \delta n$ drop out and so do the terms which are proportional to $\!\not\!\delta c$ and are not in the total derivative (those terms will contribute to the corner symplectic potential). The latter is due to the divergenceless of the Lovelock tensor, from which one derive the relation $D_b \pi^b_a+\bar\sigma^{bc}{}_{a}K_{bc}=0$
\begin{equation}
\begin{split}
\label{eq.boundaryterm2}
    & n_\mu \Theta_1^\mu=-4\pi^i_j\delta K^j_i-2(\pi^{cb}K^a_c-D_c\bar\sigma^{bac}) \delta q_{ab}-2D_c(\bar\sigma^{bac}\delta q_{ba}+\pi^{cb}\!\not\!\delta c_b).
\end{split}
\end{equation}

Let us notice that only the fully-projected Riemann tensor enter in $\pi^\alpha_\beta$ because, otherwise, there will be at least a second $n$ vector (with either upper or lower index) and the contribution would vanish by symmetry. Therefore, pull-backing it, one proves that
\begin{equation}
\label{eq.pi}
 \pi^i_j=\frac{k}{2^k}\delta^{ia_2b_2a_3b_3\dots a_{k}b_{k}}_{jc_2d_2c_3d_3\dots c_{k}d_{k}}\prod_{i=2}^{k} \pqty{R^{c_i d_i}_{a_i b_i}-2 K^{c_i}_{a_i}  K^{d_i}_{b_i}}=\frac{1}{2}p^{iab}_{jcd}(1)\pqty{R^{c d}_{a b}-2 K^{c}_{a}  K^{d}_{b}}.
\end{equation}
This relation ensures the cancellation between the $\delta K$ terms from the sum $\delta L_1+n_\mu\Theta_1^\mu$. The other terms are either (i) proportional to $\delta q$ (and thus contribute to the generalized Brown-York tensor) or (ii) inside a total boundary derivative (and thus contribute to the corner symplectic potential). While this separation is clear in \eqref{eq.boundaryterm2}, it is less so in \eqref{eq.boundaryterm1}. If we decompose the variation of the intrinsic curvature as $\delta q^{ce}R_e{}^d{}_{ab}+ q^{ce}\delta R_e{}^d{}_{ab}$, we see that the first term falls in case (i). We focus on the second term, that we rewrite as
\begin{equation}
    p^{iab}_{jcd}(s)q^{ce}\delta R_e{}^d{}_{ab}=2p^{iab}_{jcd}(s)q^{ce}q^{df}D_b D_e\delta q_{af}
\end{equation}
thanks to the antisymmetry of $p(s)$. Integrating by parts once produces a term in (ii) and\footnote{The trick to show that the l.h.s can be integrated exactly might have become familiar by now.}
\begin{equation}
\label{eq.boundaryterm3}
    -4(k-1)\int_0^1\dd s\, D_b(p^{iab}_{jcd}(s) K^j_i)q^{ce}q^{df} D_e\delta q_{af}=-4(k-1)p^{iab}_{jcd}(1) D_b K^j_i q^{ce}q^{df} D_e\delta q_{af}.
\end{equation}
Following a reasoning analogous to that leading to \eqref{eq.pi}, we can use Gauss-Codazzi relation for a single contraction of the Riemann tensor with $n$ to show that
\begin{equation}
    \bar\sigma^a{}_{cd}=2(k-1)p^{iab}_{jcd}(1)D_b K^j_i,
\end{equation}
so that the previous result \eqref{eq.boundaryterm3} is just $2\bar\sigma^{abc}D_c\delta q_{ab}$.
Integrating by parts again, we get another term in (i) and another in (ii).\footnote{Notice that these have the correct form to cancel the corresponding $\bar\sigma$ terms in the symplectic potential.}~Finally, let us collect the (i) terms in the generalized Brown-York tensor
\begin{equation}
    t_{ab}=4\pi^c_{(a} K_{b)c}
    -q_{ab} L_1-4(k-1)\int_0^1\dd s\, p^{ice}_{jd{(a}}(s) K^j_i R_{b)}{}^d{}_{ce},
\end{equation}
and the (ii) terms in the corner symplectic potential
\begin{equation}
    \Theta_2^a=
    -2\pi^{ab}\!\not\!\delta c_b+4(k-1)\int_0^1\dd s\, p^{iab}_{jcd}(s) K^j_i q^{ce}q^{df}D_e\delta q_{bf}.
\end{equation}

\bibliographystyle{JHEP}
\bibliography{References}
\end{document}